\documentclass[usenatbib]{mn2e}
\usepackage{apjfonts,amsfonts,amsmath,amssymb,ctable,verbatim,xcolor}
\usepackage{physics}

\newcommand{\gizmourl}{\href{http://www.tapir.caltech.edu/~phopkins/Site/GIZMO.html}{\url{http://www.tapir.caltech.edu/~phopkins/Site/GIZMO.html}}}
\newcommand{\FIREurl}{\href{http://fire.northwestern.edu}{\url{http://fire.northwestern.edu}}}

\newcommand{\acknowledgments}[1]{\begin{small}\section*{Acknowledgments}\end{small}{\noindent #1}\vspace{5pt}}
\newcommand{\datastatement}[1]{\begin{small}\section*{Data Availability Statement}\end{small}{\noindent #1}\vspace{5pt}}
\defcitealias{GrudicHopkins2018}{G18}
\newcommand{\Alf}{{Alfv\'en}}

\title[Rapid BH Growth in Dense Clouds]{Hyper-Eddington Black Hole Growth in Star-Forming Molecular Clouds and Galactic Nuclei: Can It Happen?}

\author[Y.\ Shi et al.]{
\vspace{0.2cm}
\parbox[t]{\textwidth}{
Yanlong Shi,$^{1}$\thanks{E-mail: yanlong@caltech.edu}
Kyle Kremer,$^{1,2}$
Michael Y. Grudi\'c,$^{2}$
Hannalore J. Gerling-Dunsmore,$^{3,4}$
Philip F. Hopkins$^{1}$
} \vspace*{4pt} \\
$^{1}$TAPIR, Mailcode 350-17, California Institute of Technology, Pasadena, CA 91125, USA\\
$^{2}$The Observatories of the Carnegie Institution for Science, Pasadena, CA 91101, USA\\
$^{3}$ Department of Astrophysical and Planetary Sciences, University of Colorado, 391 UCB, Boulder, CO 80309-0391, USA\\
$^{4}$JILA, University of Colorado and National Institute of Standards and Technology, 440 UCB, Boulder, CO 80309-0440, USA
}

\date{}
\begin{document}
\maketitle

\begin{abstract}
Formation of supermassive black holes (BHs) remains a theoretical challenge. In many models, especially beginning from stellar relic ``seeds,'' this requires sustained super-Eddington accretion. While studies have shown BHs can violate the Eddington limit on accretion disk scales given sufficient ``fueling'' from larger scales, what remains unclear is whether or not BHs can actually capture sufficient gas from their surrounding ISM. We explore this in a suite of multi-physics high-resolution simulations of BH growth in magnetized, star-forming dense gas complexes including dynamical stellar feedback from radiation, stellar mass-loss, and supernovae, exploring populations of seeds with masses $\sim 1-10^{4}\,M_{\odot}$. In this initial study, we neglect feedback from the BHs: so this sets a strong upper limit to the accretion rates seeds can sustain. We show that stellar feedback plays a key role. Complexes with gravitational pressure/surface density below $\sim 10^{3}\,M_{\odot}\,{\rm pc^{-2}}$ are disrupted with low star formation efficiencies so provide poor environments for BH growth. But in denser cloud complexes, early stellar feedback does not rapidly destroy the clouds but does generate strong shocks and dense clumps, allowing $\sim 1\%$ of randomly-initialized seeds to encounter a dense clump with low relative velocity and produce runaway, hyper-Eddington accretion (growing by orders of magnitude). Remarkably, mass growth under these conditions is almost independent of initial BH mass, allowing rapid IMBH formation even for stellar-mass seeds. This defines a necessary (but perhaps not sufficient) set of criteria for runaway BH growth: we provide analytic estimates for the probability of runaway growth under different ISM conditions.
\end{abstract}

\vspace{-0.5cm}
\begin{keywords}
black hole physics -- accretion, accretion discs -- quasars: supermassive black holes -- galaxies: star formation
\end{keywords}

\vspace{-0.3cm}
\section{Introduction}
\label{sec:intro}

Observations have demonstrated the existence of supermassive black holes (BHs) with masses $M_{\rm bh} \sim 10^9 M_\odot$ in quasars at very high redshift ($z \gtrsim 7$) when the Universe was less than a billion years old  \cite[e.g.,][]{FanNarayanan2001,WangYang2021}, which implies that these BHs must accrete rapidly from their ``seeds'' \citep{InayoshiVisbal2020}. The physical origin of these seeds remains deeply uncertain, but popular models including direct collapse of super-massive stars with masses $\sim 10^{4}-10^{6}\,M_{\odot}$ \cite[e.g.,][]{BegelmanVolonteri2006,ReganVisbal2017,CorbettMoranGrudic2018,ChonOmukai2020}, runaway mergers in globular clusters \cite[e.g.,][]{PortegiesZwartBaumgardt2004,2020ApJ...891...94B,2020MNRAS.493.2352A,KremerSpera2020,RizzutoNaab2021,ShiGrudic2021,FragioneKocsis2021,DasSchleicher2021}, remnants from Population III stars  \cite[e.g.,][]{MadauRees2001,RyuTanaka2016}, and relics of ``standard'' stellar evolution (e.g.\ Population II) stars generally produce seeds with masses $\ll 10^{4}\,M_{\odot}$. Given that the $e$-folding time of a BH growing at the Eddington limit\footnote{Throughout, we will follow standard convention and define the Eddington {\em luminosity} as the usual $L_{\rm Edd} = 3.2\times10^{4}\,L_{\odot}\,(M_{\rm bh}/M_{\odot})$, and the ``Eddington mass-accretion rate'' as the accretion rate which would produce $L_{\rm Edd}$ given a canonical reference radiative efficiency $\epsilon_{r} = 0.1$ ($L = \epsilon_{r}\,\dot{M}\,c^{2}$), so $\dot{M}_{\rm Edd} \approx M_{\rm bh}/(45\,{\rm Myr})$.} with a canonical radiative efficiency of $\sim 0.1$ is $\sim 50\,$Myr, almost all of these models require a sustained period of super or hyper-Eddington accretion in the early Universe to be viable \cite[e.g.,][]{PezzulliValiante2016}. This is especially important at masses $\ll 10^{5}\,M_{\odot}$, as various studies have shown that once larger ``super-massive'' mass scales are reached, the gravity of the BH can capture gas from larger radii and lead to runaway growth \citep{LiHernquist2007,DiMatteoColberg2008,2012MNRAS.424.1461L,2013ApJ...771..116J,WeinbergerSpringel2018,2018MNRAS.478.5063H,ZhuLi2020,2021ApJ...917...53A}. But unless one invokes exotic formation mechanisms, a sustained rapid accretion phase is necessary to grow BHs from the stellar ($\sim 10-100\,M_{\odot}$) to super-massive ($\gg 10^{4}\,M_{\odot}$) mass scale \citep{2012MNRAS.424.1461L,2016MNRAS.457.3356V}

There is a well-established and rapidly-growing body of work demonstrating that compact objects can, in fact, exceed the naive ``Eddington accretion rate'' $\dot{M}_{\rm Edd}$ by large factors (up to $\gtrsim 1000$) on scales of the accretion disk itself (recently, see e.g.\ theoretical arguments by \citealt{2016MNRAS.459.3738I,2019ApJ...880...67J,2020ApJ...905...92P,2021PASJ...73..450K,2022PASJ..tmp...17B}, empirical arguments in \citealt{2021A&A...645A..78B,2022MNRAS.509.3599T}, or for reviews, \citealt{PezzulliValiante2016,2019ffbh.book..195M,2019ConPh..60..111S,2019BAAS...51c.352B} and references therein). But these studies generally assume a constant hyper-Eddington ($\sim 10^{3}\,\dot{M}_{\rm Edd}$) influx of gas from larger scales onto the accretion disk as their ``outer boundary condition.'' What remains deeply unclear is whether or not a seed BH -- especially at stellar mass scales -- could actually capture gas from the interstellar medium at a sufficient rate to sustain this accretion, and for long enough that the total mass supplied would be able to grow the BH by many $e$-foldings. There has been some theoretical work on the topic, but it has generally either considered idealized models where the gas around the seed sits in a common potential well and accretes instead of being multi-phase and turbulent, rapidly forming stars \citep[see e.g.][]{2020MNRAS.497..302T,2022arXiv220106584P}, or considered only galactic ($\gg$\,pc) scales \citep[e.g.][]{2022arXiv220108766M} where  especially with BHs already $\gg 10^{4}\,M_{\odot}$, sustaining super-Eddington inflow to a nuclear region at least appears viable \citep{2018MNRAS.478.5063H,ReganDownes2019,ZhuLi2020,2021ApJ...917...53A}. 

The problem is that in the realistic ISM, order-of-magnitude estimates  such as those in \citet{2013ApJ...771..116J} suggest that the rate of gravitational capture of gas from the surrounding ISM -- the Bondi-Hoyle rate \citep{HoyleLyttleton1939,Bondi1952} -- should be extremely small unless the seed is already super-massive. Consider the standard expression
\begin{align}
    \dot{M}_{\rm Bondi} \approx \frac{4\pi\,G^2\,M_{\rm bh}^2\rho}{(c_{\rm s}^2+\delta V^2)^{3/2}}.
    \label{equ:bondi-hoyle-rate}
\end{align}
where $\rho$, $c_{\rm s}$, and $\delta V$ are the density, sound speed, and gas-BH relative velocity. In the diffuse/warm ISM, this gives $\dot{M}_{\rm Bondi}/\dot{M}_{\rm Edd} \sim 10^{-6}\,(M_{\rm bh}/10\,M_{\odot})\,(n/{\rm cm^{-3}})$ -- vastly sub-Eddington. In dense ($n \gtrsim 100\,{\rm cm^{-3}}$) cold molecular gas (sound speed $\sim 0.1\,{\rm km\,s^{-1}}$), $\dot{M}_{\rm Bondi}$ would be much larger {\em if the gas were laminar and the BH stationary} -- this is akin to the idealized non-turbulent models above. The problem is that realistic cold molecular gas in the ISM is clumpy and dynamical and turbulent, with star formation and stellar feedback generating large random motions -- i.e.\ large $\delta V$ \citep{larson:gmc.scalings,goodman:1998.lws.dependence.on.tracers,evans:1999.sf.gmc.review,stanimirovic:1999.smc.hi.pwrspectrum,elmegreen:2004.obs.ism.turb.review}. As we show below, assuming relative velocities are of order typical gravitational/virial velocities in the cloud then gives $\dot{M}_{\rm Bondi}/\dot{M}_{\rm Edd} \sim 10^{-4}\,(\langle n_{\rm cl} \rangle / 100\,{\rm cm^{-3}})^{1/2}\,(M_{\rm bh}/10\,M_{\odot})\,(10^{6}\,M_{\odot}/M_{\rm cl})$ -- once again, vastly sub-Eddington. Previous analytic and simulation models of this ``turbulent Bondi-Hoyle problem'' in idealized driven turbulence have argued that vorticity and turbulent magnetic fields will suppress the {\em average} accretion rates even relative to this (pessimistic) result \citep{KrumholzMcKee2006,BurleighMcKee2017}.

However, it is also clear from many studies of star formation that turbulence in dense gas also promotes the existence of extremely dense shocks and clumps in the gas \citep[see e.g.][]{klessen:2000.pdf.supersonic.turb,elmegreen:sf.review,vazquez-semadeni:2003.turb.reg.sfr,MacLowKlessen2004,federrath:2008.density.pdf.vs.forcingtype,goodman:2009.dendrogram.sims,federrath:2010.obs.vs.sim.turb.compare,hopkins:2012.intermittent.turb.density.pdfs,squire.hopkins:turb.density.pdf}, which can have low internal velocity dispersions and play a crucial role in turbulent fragmentation and star formation \citep{mckee:2007.sf.theory.review,hennebelle:2008.imf.presschechter,hopkins:excursion.ism,hopkins:excursion.imf,hopkins:excursion.clustering,hopkins:excursion.imf.variation,hopkins:frag.theory,guszejnov:gmc.to.protostar.semi.analytic,murray:2017.turb.collapse}. So it is possible that a more realistic model might allow for hyper-Eddington accretion in rare (but not impossible) cases in these environments. In this study, we therefore extend the series of simulations of dense, star forming environments used previously to study star and star cluster formation in \cite{GrudicHopkins2018,guszejnov:2018.isothermal.nocutoff,GrudicGuszejnov2018,grudic:max.surface.density,grudic:sfe.gmcs.vs.obs,grudic:2019.imf.sampling.fx.on.gmc.destruction,ShiGrudic2021}, to explore BH seed growth in dynamic, star-forming environments akin to dense giant molecular clouds (GMCs) and galactic nuclei. 

In this first study, we neglect feedback from the accreting BHs themselves. This is obviously a major simplification, especially for BHs accreting above the Eddington limit -- however, the form and strength of feedback from BHs in this regime remains highly uncertain (see references above), and we wish to identify whether or not sustaining hyper-Eddington accretion is even remotely possible on these scales. Clearly, accretion {\em without} BH feedback represents a relatively strong upper limit to the maximum possible BH seed growth. We can then use the conditions identified here as necessary for such accretion to run simulations including BH feedback, with various parameterizations.

In \S~\ref{sec:simulations}, we describe our simulation methods. Then in \S~\ref{sec:results} we present results, including BH mass evolution in different clouds and its dependence on different initial conditions (ICs). In \S~\ref{sec:discussions}, we analyze the effects of different physics and simulation ICs, give simple analytic formulae for  the conditions required for runaway accretion, and discuss some major caveats of our work in \S~\ref{sec:caveats}. Finally, we conclude in \S~\ref{sec:conclusions}.

\begin{footnotesize}
\ctable[caption={{\normalsize Initial conditions (ICs) of our ``fiducial'' reference simulations. Here we show three groups of simulations with low, medium, and high initial mean surface density ($\Bar{\Sigma}_0$). In each group, the clouds have radii ($R_{\rm cl}$) of 5, 50, 500 pc.. Subsequent columns give the approximate initial total cloud mass ($M_{\rm cl}$), initial free-fall time ($t_{\rm ff,\,c}$), gas cell mass/resolution ($m_{\rm gas}$), Plummer-equivalent force softening for star particles ($\epsilon_{\rm soft}^{\rm star}$), and for BHs ($\epsilon_{\rm soft}^{\rm bh}$), and additional notes.}}\label{tab:clouds},center,star,
]{llllllll}{
}{
\toprule
$\Bar{\Sigma}_0$ [$M_{\odot}$/pc$^2$] & $R_{\rm cl}$ [pc] & $M_{\rm cl}$ [$M_{\odot}$] &  $t_{\rm ff,\,c}$ [Myr] & $m_{\rm gas}$ [$M_{\odot}$] & $\epsilon_{\rm soft}^{\rm star}$ [pc] & $\epsilon_{\rm soft}^{\rm bh}$ [pc]  & Notes \\
\midrule
130 & 5 & $10^{4}$ & 2 & 0.005 & 0.04 & 0.04 & \\
130 & 5 & $10^{4}$ & 2 & 0.04 & 0.09 & 0.09 & No-feedback (low-resolution) variant\\
130 & 50 & $10^{6}$ & 6 & 0.5 & 0.21 & 0.21 & \\
130 & 500 & $10^{8}$ & 20 & 50 & 0.96 & 0.31 & \\
\midrule
1300 & 5 & $10^{5}$ & 0.6 & 0.4 & 0.19 & 0.19 & \\
1300 & 50 & $10^{7}$ & 2 & 40 & 0.89 & 0.31 & \\
1300 & 500 & $10^{9}$ & 6 & 500 & 2.06 & 0.31 & \\
\midrule
13000 & 5 & $10^{6}$ & 0.2 & 0.5 & 0.21 & 0.21 & \\
13000 & 5 & $10^{6}$ & 0.2 & 4 & 0.41 & 0.31 & Varied metallicity test series\\
13000 & 50 & $10^{8}$ & 0.6 & 6 & 0.48 & 0.31 & Highest resolution; $M_{\rm bh}\in(10, 100)\,M_\odot$\\
13000 & 50 & $10^{8}$ & 0.6 & 50 & 0.96 & 0.31 & \\
13000 & 50 & $10^{8}$ & 0.6 & 400 & 1.91 & 0.31 & Varied BH seed number test series\\
13000 & 500 & $10^{10}$ & 2 & 40000 & 8.89 & 0.31 & \\
\bottomrule
}
\end{footnotesize}

\begin{figure*}
    \centering
    \includegraphics[width=\linewidth]{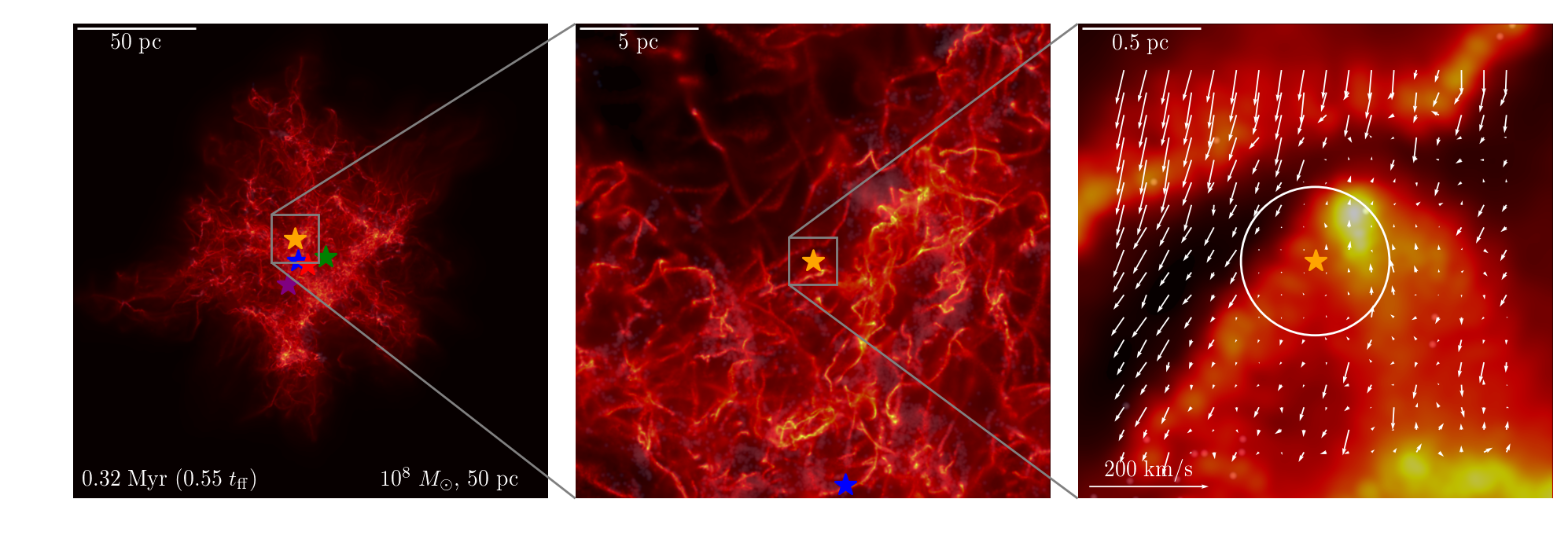}
    \vspace{-20 pt}
    \caption{An example visualization of the simulations at different scales. The simulation starts with a massive dense gas complex of $M_{\rm cl}=10^8~M_{\odot}$, $R_{\rm cl}=50~{\rm pc}$, and $256^3$ resolution, and we choose the snapshot where the seeds have the highest accretion rate. Here we show the distribution of the gas surface density as the black-red-orange color map. As stars, we show the BH seeds that undergo most significant growth. \textbf{Left}: Visualization of the whole complex; we see the BHs that accrete most are located at the dense region near the center of the system. \textbf{Middle}: 10 times zoom-in centered on the BH that shows the most significant mass growth at this time. \textbf{Right}: Further zoom-in of the region near the BH and the velocity field of gas (vectors, length proportional to magnitude of $|\mathbf{v}_{\rm bh}-\mathbf{v}_{\rm gas}|$) in the vicinity of the BH. Here the circle denotes the sink radius. We see a dense clump intersecting the sink radius, with a the relative velocity that is quite small near the BH.}
    \label{fig:visualization}
\end{figure*}

\section{Simulations}
\label{sec:simulations}

Our simulation numerical methods are identical to those described and tested fully in \citet{GrudicHopkins2018,GrudicGuszejnov2018,HopkinsWetzel2018,GrudicGuszejnov2021,GrudicKruijssen2021}, modulo the addition of BH seeds described below, so we briefly summarize here. We use the code \textsc{GIZMO}\footnote{A public version of {\small GIZMO} is available at \gizmourl} \cite[]{Hopkins2015} in Meshless Finite Mass (MFM) mode, with magnetohydrynamics (MHD) solved as in \citet{hopkins:mhd.gizmo,hopkins:cg.mhd.gizmo}, self-gravity with adaptive Lagrangian force-softening, radiative cooling from $1-10^{10}$\,K, including molecular, metal-line, fine-structure, photo-electric, ionization and other processes as well as star formation in dense, locally-self-gravitating gas \citep{hopkins:virial.sf,GrudicHopkins2018}, and stellar feedback following the FIRE-2 implementation of the Feedback In Realistic Environments (FIRE\footnote{\FIREurl}) physics \citep{HopkinsWetzel2018,hopkins:fire3.methods}. In these models ``star particles'' each repreent IMF-averaged ensembles of stars (rather than resolving individual stars and proto-stars as in \citealt{grudic:starforge.methods,guszejnov:2020.starforge.jets}), which evolve along standard stellar evolution models to return mass, metals, momentum, and energy to the ISM in the form of supernovae and O/B and AGB winds \citep{hopkins:sne.methods} as well as acting on the gas via radiative heating, photo-ionization, and radiation pressure \citep{hopkins:radiation.methods}. Simulations with these methods have been previously used to study many properties of GMCs, galactic nuclei, and star clusters, including their observed star formation efficiencies, cluster dynamics and mass profiles, young massive cluster internal structure, globular cluster demographics, and gas emission properties \citep[see references above and e.g.][]{GrudicHopkins2018,GrudicGuszejnov2018,GrudicGuszejnov2021,GrudicKruijssen2021,FukushimaYajima2021}. 

We extend these simulations by adding a population of ``seed'' BHs (sink particles) to the ICs, which can accrete gas from the surrounding medium, but otherwise feel only gravitational dynamics (we do not model BH feedback or BH-BH mergers).

\subsection{Black Hole Accretion}
\label{sec:simulations:bh-accretion}

Our BH seeds/sink particle prescription is a simplified version of that numerically presented in \citet{grudic:starforge.methods}. Gas is accreted onto a sink if it meets three criteria:
\begin{enumerate}
\item It is within the sink radius $r_{\rm sink}$ of the BH: $r=|\mathbf{r}_{\rm gas}-\mathbf{r}_{\rm bh}|< r_{\rm sink}$. 
\item It is bound to the BH, including kinetic, thermal, and magnetic support: $u_{\rm thermal} + (1/2)\,v_{\rm A}^{2} + (1/2)\,\delta V^{2} < G\,M_{\rm sink}/r$, where $u_{\rm thermal}$ is the specific thermal energy, $v_{\rm A}$ the \Alf\ speed, and $\delta V^{2} \equiv |\mathbf{v}_{\rm gas} - \mathbf{v}_{\rm bh}|^{2}$.
\item Its angular momentum is insufficient to support a circular orbit with radius larger than $r_{\rm sink}$ \cite[]{BateBonnell1995}, i.e.\ $j_{\rm gas} <\sqrt{G\,M_{\rm sink}\,r_{\rm sink}}$ where $j_{\rm gas}$ is the specific angular momentum of the gas cell (evaluated at its center-of-mass location).
\end{enumerate}
If a gas cell somehow meets all these criteria with two BHs simultaneously, it will accrete onto whichever is closer. 

We must choose $r_{\rm sink}$ in each simulation. This is usually set to something like the simulation resolution (typical inter-cell separation $\delta r$), and would ideally resolve the Bondi radius, $R_{\rm Bondi} \sim G\,M_{\rm bh}/(c_{\rm s}^{2} + \delta V^{2})$, i.e.\ $r_{\rm sink} \sim R_{\rm Bondi} \gtrsim \delta r$. But in our Lagrangian, dynamical simulations (1) the spatial resolution is not fixed, but scales as $\delta r \sim (\rho/m_{\rm gas})^{1/3}$, and (2) the Bondi radius fluctuates dramatically (as we will show), and varies between seeds. In the ``worst case'' scenario, assume accretion is coming from the low-density diffuse intra-cloud medium (density $\rho \sim \langle \rho \rangle \sim 3\,M_{\rm cl}/4\pi\,R_{\rm cl}^{3}$) with virial or free-fall level relative velocities $\delta V \sim v_{\rm cl} \sim (G\,M_{\rm cl}/R_{\rm cl})^{1/2} \gg c_{s}$. This would give $R_{\rm Bondi} \sim (M_{\rm bh}/M_{\rm cl})\,R_{\rm cl}$, so resolving the Bondi radius ($\delta r \lesssim R_{\rm Bondi}$ in the same diffuse mean-density gas) would require a prohibitive number of cells $N_{\rm cells} \sim (R_{\rm cl}/\delta r)^{3} \gtrsim (M_{\rm cl}/M_{\rm bh})^{3}$. However, as we noted above and will show more rigorously below, the accretion rates from such diffuse gas are orders-of-magnitude below Eddington, and (even if well-resolved) would contribute essentially nothing to the total BH accretion in our simulations. Therefore consider instead the ``best-case'' scenario for accretion: since the turbulence in the molecular clouds has rms Mach numbers $\mathcal{M}_{\rm cl} \sim v_{\rm cl}/c_{s} \sim 10-100$, radiative shocks can produce regions with very high densities $\rho \sim \langle \rho \rangle\,\mathcal{M}_{\rm cl}^{2}$, and low relative velocities $\delta V \lesssim c_{s}$ \citep{Vazquez-Semadeni1994,PadoanNordlund1997,MacLowKlessen2004}. Under these conditions, the Bondi radii will be well-resolved ($\delta r \lesssim R_{\rm Bondi}$) so long as $N \gtrsim \mathcal{M}^{-8}\,(M_{\rm cl}/M_{\rm bh})^{3}$ -- a huge relief ($\propto \mathcal{M}^{8}$) in resolution requirements (which would be easily satisfied by every simulation in this paper). As we will show, regions akin to this idealized example dominate the actual accretion in the simulations.

In practice, we choose a sink radius by estimating a ``characteristic'' Bondi radius $b_{\rm c}$ by assuming $M_{\rm bh,\,c}=100\,M_\odot$, and considering two limits: $\delta V \lesssim c_{\rm s}$ (assuming a mean temperature of $100\,$K, typical in our simulations) so {$b_1 = G\,M_{\rm bh,\,c}/c_{\rm s}^{2}$},  and $\delta V \sim v_{\rm cl} \gg c_{\rm s}$ so $b_2 \approx (M_{\rm bh}/M_{\rm cl}) R_{\rm cl}$, and then take $b_{\rm c}=\min (b_1, b_2)$. We have verified in post-processing that in all cases which produce ``interesting'' runaway BH growth, the Bondi radii {\em during the phase where the BH actually accretes rapidly} is at least marginally resolved, as expected from the argument above.

{We wish to reminder the readers again that the mass ``accreted'' in the simulation is not the actual mass swallowed by BHs due to multiple feedback effects (for details see Sec.~\ref{sec:caveats}). The sink radius $r_{\rm sink}$ is the actual resolution limit for BH accretion, while the physics from $r_{\rm sink}$ to the Schwarzchild radius is not resolved in this simulation, but it does not impact the science goal of this article. For completeness, an estimate considering BH radiative feedback from the previous analytic work \cite[]{2016MNRAS.459.3738I} is included in Sec.~\ref{sec:discussions:hyper-Eddington}.
}

\subsection{Initial Conditions}
\label{sec:simulations:ics-setups}

We sample spherical, turbulent, and non-rotating molecular clouds or cloud complexes with different initial mean surface density ($\bar{\Sigma}_0 \equiv M_{\rm cl}/\pi\,R_{\rm cl}^{2} \approx 100, 10^3, 10^4\,M_\odot/{\rm pc}^2$) and initial radius ($R_{\rm cl}=5, 50, 500\,{\rm pc}$) following the setup and results of \citetalias{GrudicHopkins2018}, where each group with the same surface density was shown to have similar star formation efficiency. Note that these parameters are motivated by massive, dense star-forming cloud and ``clump'' complexes seen in high-redshift galaxies and starburst galaxy nuclei, with only the smaller and lowest-$\bar{\Sigma}_{0}$ clouds analogous to massive GMCs in the Milky Way. Each initial cloud is uniformly magnetized, we also set $E_{\rm turb}/|E_{\rm grav}|=1$ and $E_{\rm mag}/E_{\rm grav}=0.1$, where $E_{\rm turb}$, $E_{\rm mag}$, and $E_{\rm grav}$ are the turbulence (kinetic) energy, magnetic field density, and gravitational binding energy respectively. The clouds serve as the mass reservoirs for BH accretion.

We then insert an ensemble of BH seeds into the IC. Typically, for every seed, the mass ranges within $1~M_\odot \le M_{\rm bh} \le 10^4~M_{\odot}$ and are uniformly distributed in $\log M_{\rm bh}$. The initial position of seeds are sampled randomly but statistically uniformly within the cloud. The initial velocity is sampled such that in each dimension it is uniformly distributed in $[-V_{\rm circ},V_{\rm circ}]$ while the total magnitude is suppressed below $V_{\rm circ}$ to ensure the seeds are bound to the cloud, where $V_{\rm circ}^2 = GM_{\rm cl}(<r)/r$ is the local circular velocity at radius $r$ (assuming uniform mass distribution). We resample seeds which would be within a small distance to the cloud ``edge'' with an outward radial velocity, since these would trivially escape without interesting dynamics.

Rather than simulating only a few BH seeds in one cloud, we include a large number of seeds in every IC so that we can sample many different seed masses and positions and kinematics. However, to avoid significant interactions among the BHs and heavy computational costs, the number of BH seed is controlled to be either below 10000, or the number such that the total BH mass does not exceed $5\%$ of the cloud initial mass $M_{\rm cl}$. For low mass clouds, we decrease the lower and upper bounds of BH seed mass sampling to ensure a sufficient number of BH seeds, which also helps ensure the Bondi radii are resolved (e.g., for $M_{\rm cl}= 10^{4}\,M_{\odot}$, $1\,M_\odot \le M_{\rm bh} \le 100 \,M_\odot$; for $M_{\rm cl}= 10^{5}\,M_{\odot}$, $10\,M_\odot \le M_{\rm bh} \le 10^3 \,M_\odot$, for $M_{\rm cl} \gtrsim 10^{6}\,M_{\odot}$, $10^{2}\,M_\odot \le M_{\rm bh} \le 10^{4} \,M_\odot$).

We use adaptive force softening to avoid divergences in our gravity evaluation or extremely small time steps. For the newly formed stars, which have the same mass as gas particles, the minimum softening length is $r_{\rm soft}^{\rm star} \sim ( m_{\rm gas}/\rho_{\rm sf})^{1/3}$, where $m_{\rm gas}$ is the mass resolution of the cloud and $\rho_{\rm sf}$ is the numerical minimum density for star formation ($1000\,{\rm cm}^{-3}$ in the simulation). For BHs, the softening radius is set as $r_{\rm soft}^{\rm bh} = \min (r_{\rm soft}^{\rm star}, b_{\rm c})$, where $b_{\rm c}$ is the characteristic Bondi-Hoyle accretion radius (introduced in \S~\ref{sec:simulations:bh-accretion}). In the simulation $r_{\rm sink}=r_{\rm soft}^{\rm bh}$, so the setup ensures the code resolves the Bondi-Holye accretion radius and the BHs interact reasonably with star particles.

For reference, we show the initial-conditions in Table~\ref{tab:clouds}. The clouds are divided into three groups with different initial mean surface density $\Bar{\Sigma}_0 = M_{\rm cl}/(\pi R_{\rm cl}^2)$, though within each group the clouds have the same set of initial radii. The fiducial resolution (number of initial gas cells) of our simulations is $128^3$, while a few low ($64^3$) and high ($256^3$) resolution runs of a subset of clouds are also included for comparison. For each fiducial simulation, the termination time scale is $2\, t_{\rm ff}$, where $t_{\rm ff}=\pi\sqrt{R_{\rm cl}^3/(8G M_{\rm cl})}$ is the initial free-fall time scale of the cloud, while for low-resolution ones the termination time is $5\,t_{\rm ff}$. Finally, BH mergers are disabled since the event rate is not significant.

\begin{figure*}
    \centering
    \includegraphics[width=\linewidth]{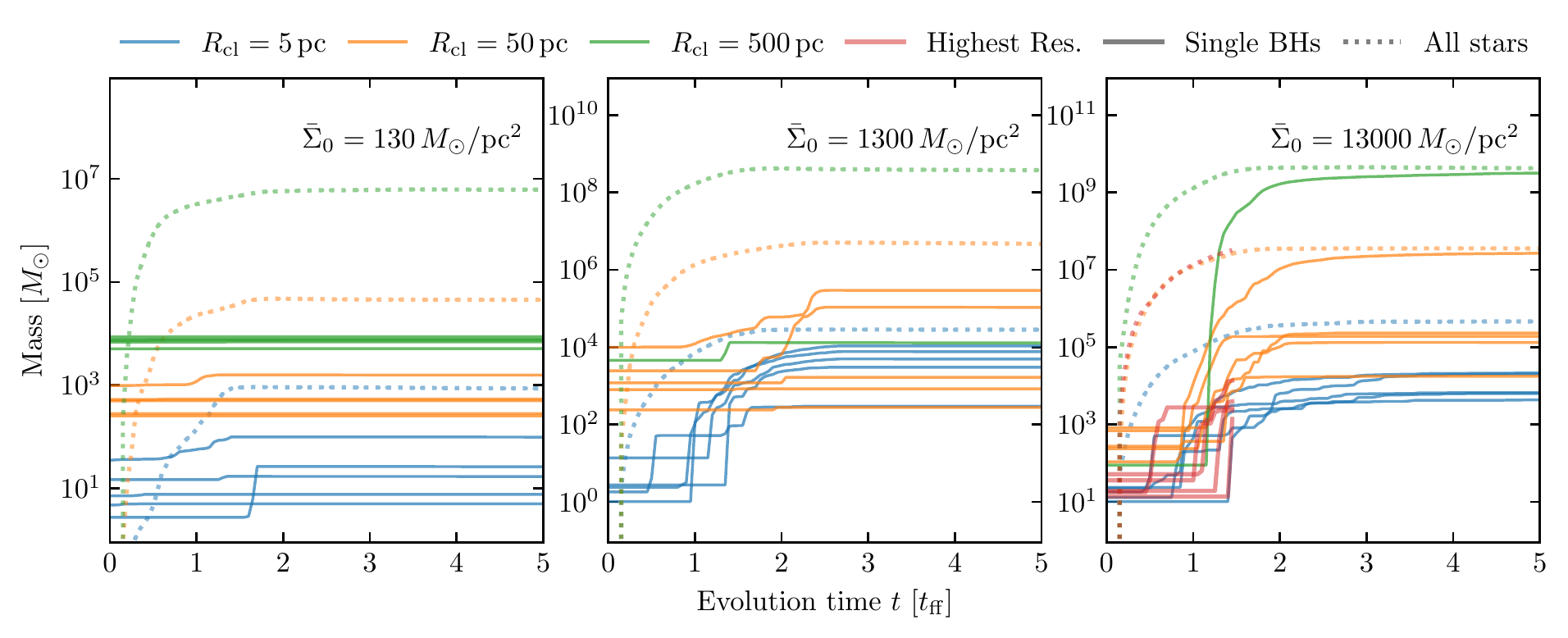}
    \vspace{-10 pt}
    \caption{Mass growth of the total mass of stars as well as (up to) five BHs that show the most significant mass growth, in the ``default'' simulations from Table~\ref{tab:clouds}, as a function of time (in units of the initial homogeneous-cloud free-fall time $t_{\rm ff}$). Here solid lines representing the mass growth of individual BHs, dotted lines representing the total mass growth of all stars summed. We show properties of complexes with different radii/masses ($r_{\rm cl}$): 5\,pc (blue), 50\,pc (orange), 500\,pc (green); and different initial mean surface density ($\Bar{\Sigma}_0$): 130\,$M_{\odot}/{\rm pc}^2$ (left), 1300\,$M_{\odot}/{\rm pc}^2$ (middle), 13000\,$M_{\odot}/{\rm pc}^2$ (right). We also show the highest-resolution run with $M_{\rm cl}=10^8\,M_\odot$ and $R_{\rm cl}=50\,{\rm pc}$ (red lines in the right panel).
    The low-density complexes feature almost no BH growth. Higher-density systems generally feature a small number of seeds which ``run away'' and can grow by orders of magnitude.
    }
    \label{fig:mass-growth}
\end{figure*}

\section{Results}
\label{sec:results}

In this section we show the major results of the simulations. As a first impression, we present the morphology of one example GMC in Fig.~\ref{fig:visualization}, which has $M_{\rm cl}=10^8\,N_\odot$, $R_{\rm cl}=50\,{\rm pc}$, and resolution of $256^3$. After $0.55\,t_{\rm ff}$ of evolution, the GMC has become quite turbulent. We also show the 5 BHs that show the most significant mass growth during the period, which are generally located near the center of the GMC. For the BH that grows most rapidly during the period (the orange star in Fig.~\ref{fig:visualization}), we show the zoomed-in distribution of gas\footnote{For illustration purposes, the color is scaled nonlinearly with the density field so as to better illustrate its morphology.} and its velocity field in the middle and right-hand panels. Near the BH's sink radius (0.313 pc), there is a dense gas clump which has very low velocity compared to the gas at the edge of the view ($\sim 50\,{\rm km/s}$); this is rapidly accreted in the time between the snapshot shown and the next simulation snapshot.

This essentially fits our expectations from Bondi-Hoyle theory, applied {\rm locally} at scales of order the Bondi radius: high gas density and low relative velocity between the BH and nearby gas create the ideal conditions for growth.

\subsection{Seed growth in different clouds}
\label{sec:growth.vs.cloud}

\begin{figure}
    \centering
    \includegraphics[width=\linewidth]{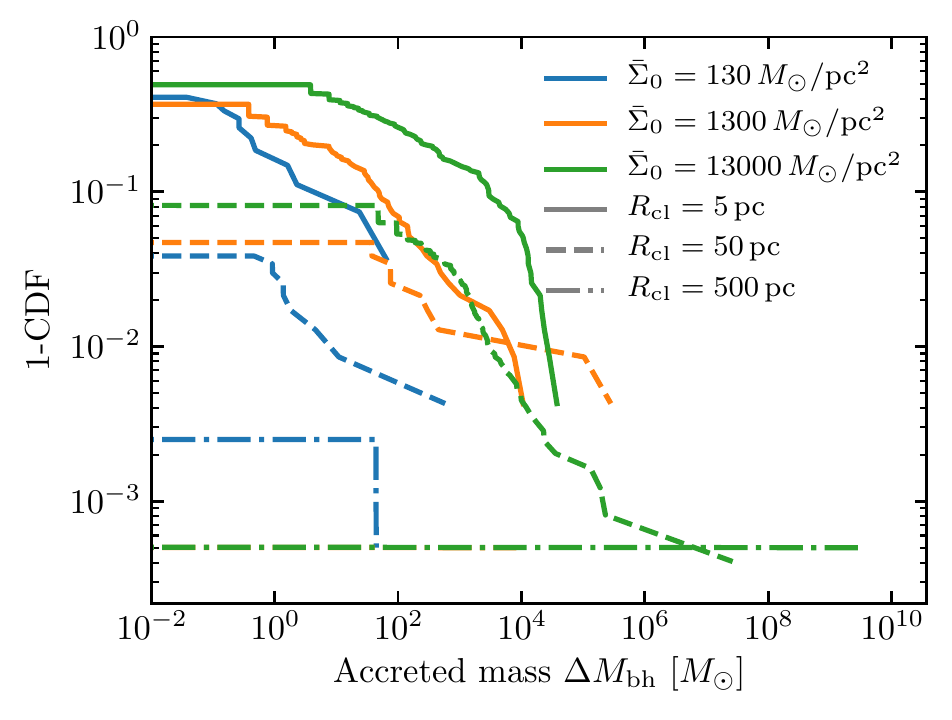}
    \vspace{-20 pt}
    \caption{Distribution of fractional mass growth in different ICs: we plot the cumulative fraction of the BH seeds which accreted some amount of mass. We denote initial surface density with color, and initial cloud radius/mass with linestyle (as labeled). Most BHs do not accrete significantly (see lower-end cutoffs). But the most high-density, massive clouds generally feature a very small number of seeds which runaway to enormous masses. }
    \label{fig:cdf}
\end{figure}

As described in the previous section, in each cloud we sampled a large number of BH seeds to study their mass growth. In Fig.~\ref{fig:mass-growth} we present the mass evolution of (up to) 5 BHs in each simulation that show the most significant mass growth. As we show below, these are {\em not} necessarily the most massive seeds in the ICs. 

For clouds with low initial surface density ($\Bar{\Sigma}_0=127\,M_{\odot}/{\rm pc^2}$) the mass growth is modest: essentially no BHs grow by more than a factor $\sim 2-3$, and in general even the most-rapidly growing only increase in mass by tens of percent. At the larger surface densities we sample, the mass of the most-rapidly-growing BHs typically increases by at least two orders of magnitude. 

Ignoring the low surface density complexes, if we consider clouds with fixed $\Bar{\Sigma}_{0}$ but different sizes $R_{\rm cl}$ (or equivalently, masses $M_{\rm cl} \equiv \pi\,\Bar{\Sigma}_{0}\,R_{\rm cl}^{2}$), we see that the final masses of the single most-rapidly-growing BHs increase with the total cloud mass, reaching as high as $\sim 3-10\%$ of the total cloud mass. Interestingly, for the lower-mass complexes, we often see several BHs exhibiting similar growth, while for the most massive complexes ($R_{\rm cl} = 500\,{\rm pc}$), one BH runs away early and then proceeds to accrete a huge amount of gas, ``stealing'' the gas supply from other seeds in the cloud. 

From the same plot, we also see that the BHs typically grow their mass quickly in a short range of time ($\Delta t \lesssim t_{\rm ff}$) starting at some time near $t \sim t_{\rm ff}$. However, for clouds with higher surface density, we see the time range becomes slightly longer; the BHs in those clouds also start to grow somewhat earlier. Moreover, as we will show below in more detail in some illustrative examples, BH growth always {\em follows} the formation of a significant mass of stars. All these features inspire us to study the effect of star formation and stellar feedback in different clouds, which is discussed in \S~\ref{sec:discussions:feeback}.

As a different way to study the probability of mass growth, we show the cumulative distribution of the final-initial mass difference for all the BHs in Fig.~\ref{fig:cdf}. For most of the BHs there is no large significant mass growth except for a small fraction ($\lesssim 10\%$) of them, which we will discuss in more detail below.

\subsection{Dependence on ICs}

\begin{figure}
    \centering
    \includegraphics[width=\linewidth]{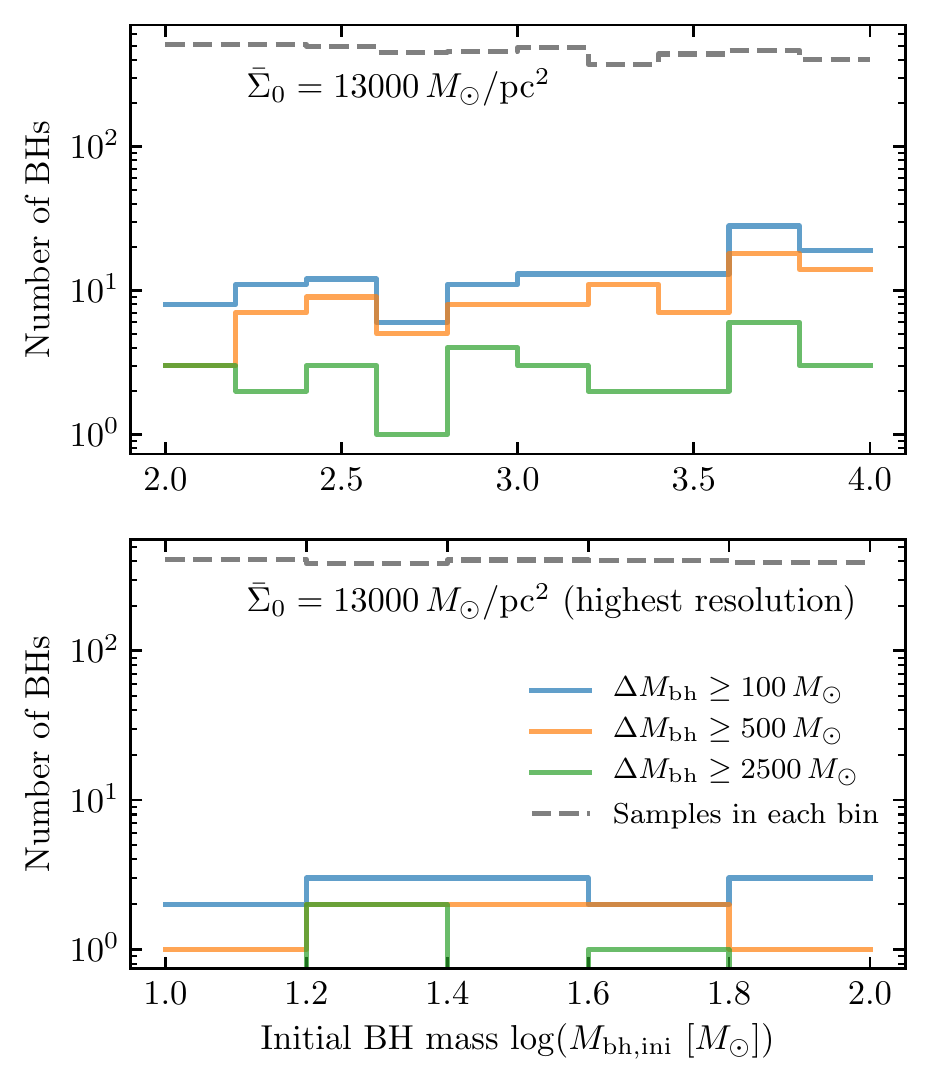}
    \vspace{-20 pt}
    \caption{Dependence of BH growth on initial seed BH mass. 
    \emph{Top}: We aggregate all seeds in the three fiducial runs ($R_{\rm cl}=5,\,50,\,500\,$pc) for our highest-density clouds, and plot both the total number of BHs in different $\sim 0.1$\,dex-wide bins of initial BH mass ($\sim 1000$ per bin, summing across the simulations), and the number in each bin which capture/accrete a mass $\Delta M_{\rm bh}$ in excess of some fixed amount. The probability of runaway accretion depends only very weakly on initial seed mass, over this range. 
    \emph{Bottom}: Same as above, but for the highest resolution cloud ($M_{\rm cl}=10^{8}\,M_\odot$, $R_{\rm cl}=50\,{\rm pc}$) which has lighter seeds (10--100 $M_\odot$). Again, there is shallow dependence on the initial seed mass.
    }
    \label{fig:initial-mass}
\end{figure}

\begin{figure*}
    \centering
    \includegraphics[width=\linewidth]{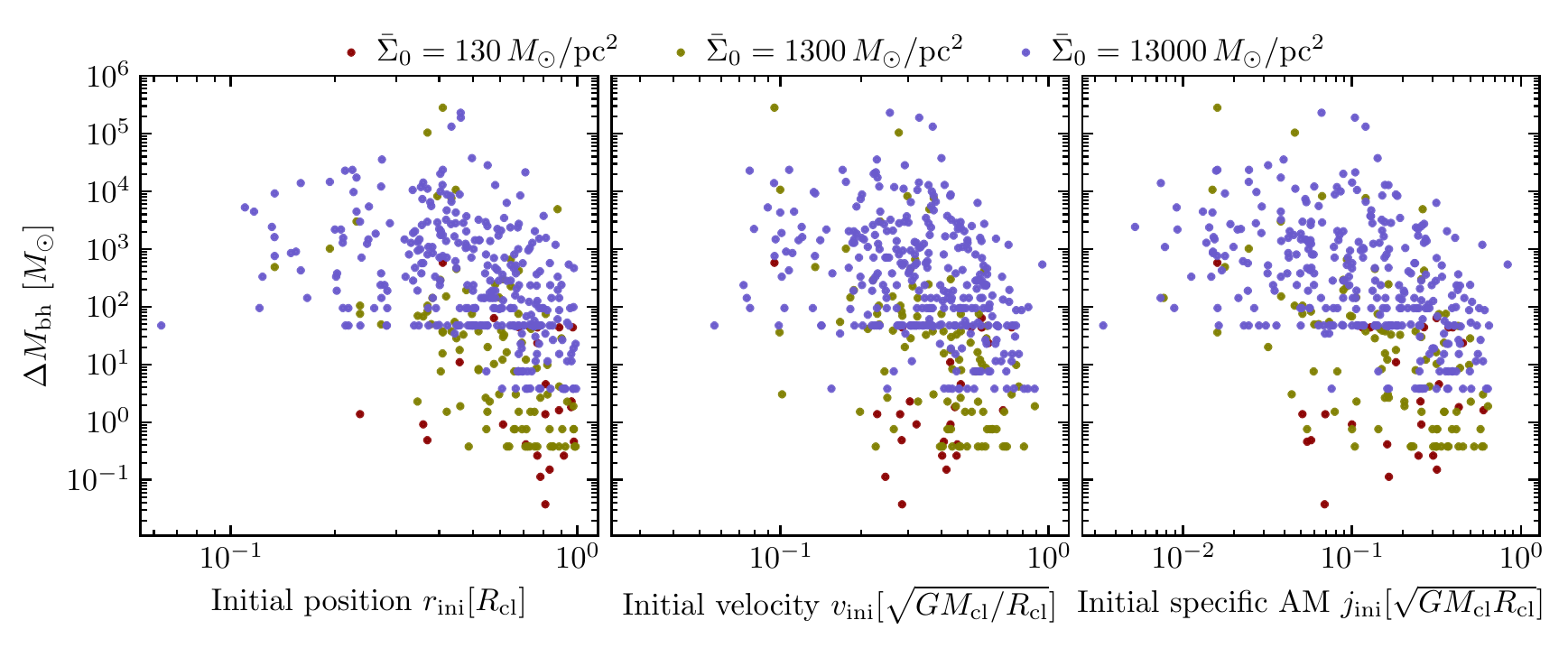}
    \vspace{-20 pt}
    \caption{Initial value of the BH seed velocity $v_{\rm ini} = |\mathbf{v}_{\rm bh}|$ (relative to the initial complex center-of-mass velocity, in units of the characteristic gravitational velocity of the complex $v_{\rm cl} = \sqrt{G\,M_{\rm cl}/R_{\rm cl}}$), initial position $r_{\rm ini} = |\mathbf{r}_{\rm bh}|$ (relative to the initial complex center-of-mass, in units of its radius $R_{\rm cl}$), {and initial value of the specific angular momentum $j_{\rm ini} = |\mathbf{r}_{\rm BH}\times \mathbf{v}_{\rm BH}|$}, for all seeds in clouds of a given initial surface density (aggregating the three different-mass/size runs). {All these quantities are compared with the amount of mass growth (final-to-initial mass difference) for each BH.} As long as the seeds do not begin too close to the ``edge'' of the cloud or with too large an initial velocity (in which case they tend to ``drift away'' rather than accrete), {or with too large specific angular momentum,} there is no strong preference for e.g.\ seeds to be at the cloud center for runaway accretion to occur. }
    \label{fig:position-velocity}
\end{figure*}

\begin{figure}
    \centering
    \includegraphics[width=\linewidth]{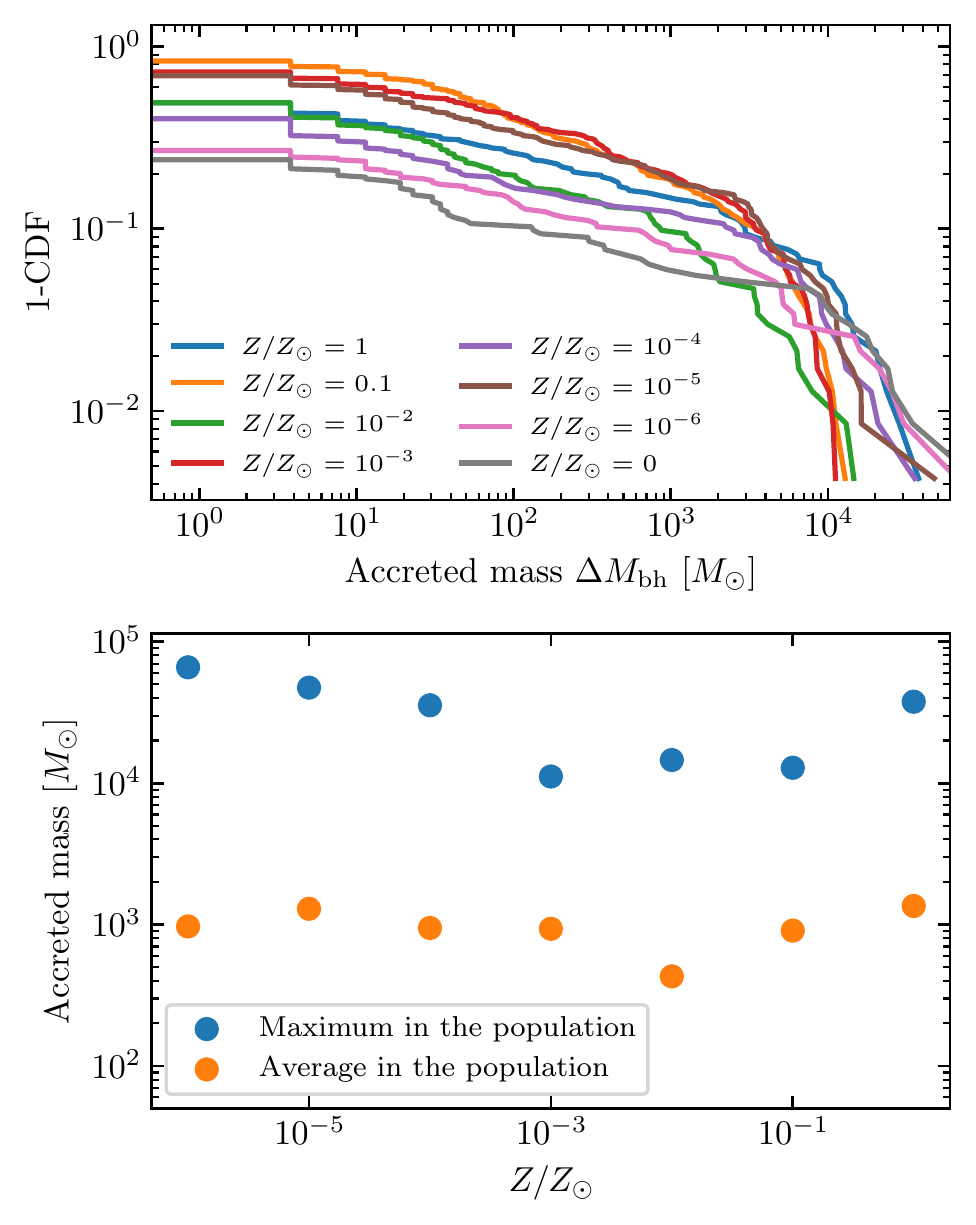}
    \vspace{-20 pt}
    \caption{Metallicity effects on accretion. We show the cumulative accreted mass distribution for all BHs as Fig.~\ref{fig:cdf}, but for a set of $M_{\rm cl}=10^6\,M_\odot$ and $R_{\rm cl}=5\,{\rm pc}$ systems simulated at low resolution systematically varying the initial metallicity. There is no strong systematic metallicity dependence.}
    \label{fig:metallicity}
\end{figure}

\begin{figure}
    \centering
    \includegraphics[width=\linewidth]{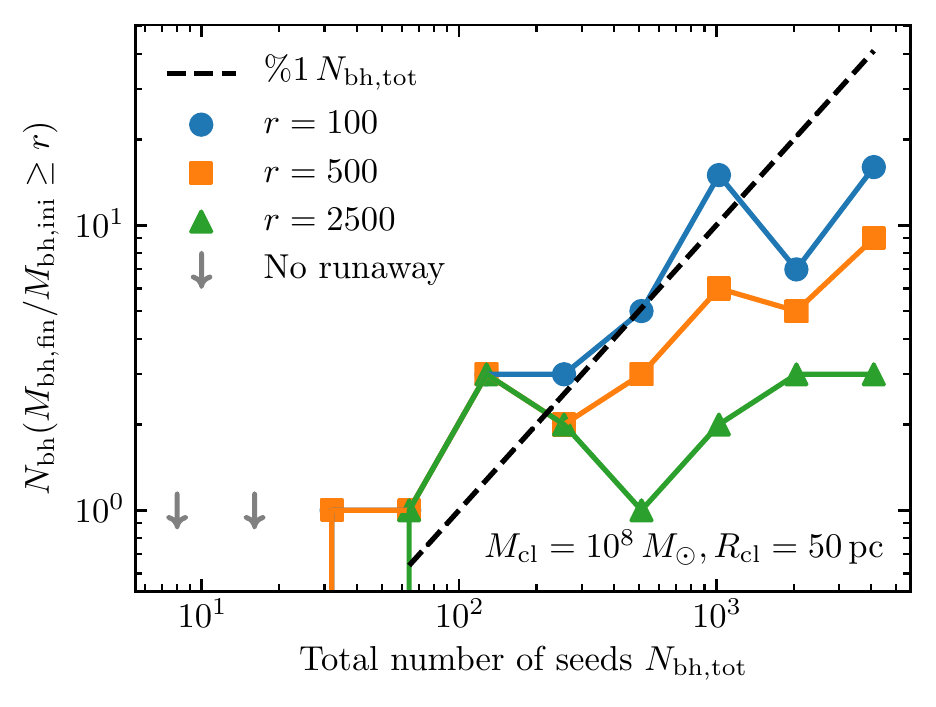}
    \vspace{-20 pt}
    \caption{Number of BHs which exceed some final-to-initial mass ratio $r$ (as a proxy for ``runaway'' growth), in simulations of an otherwise identical complex (with properties labeled) where we vary the initial seed number systematically. We require $\sim 50-100$ seeds before there is reliably at least one seed that undergoes runaway growth. Increasing the seed number further, the fraction of seeds which experience at least {\em some} significant growth is $\sim 1\%$, but we see that typically no more than one to a few BHs run away most dramatically (regardless of the seed number provided it is sufficient). }
    \label{fig:bh-number-dependence}
\end{figure}

In Fig.~\ref{fig:initial-mass} we show the dependence of mass growth ($\Delta M_{\rm bh}$) on the initial mass of the BH seeds. As we showed above, most seeds did not grow significantly. But more strikingly -- and perhaps surprisingly, given the strong dependence of the Bondi-Hoyle rate on $M_{\rm bh}$ -- we see that there is almost no correlation between the initial seed mass and BH mass growth. The particular simulation here considers seeds from $10^{2}-10^{4}\,M_{\odot}$, but we find the same (in the extended tests described below) for initial seed masses down to $\sim 10\,M_{\odot}$. 

In Fig.~\ref{fig:position-velocity}, we present the initial velocity magnitude (relative to the cloud center-of-mass), initial position, and mass growth for all BHs in the simulation. As we can see, there is no strong dependence on either the initial position or velocity magnitude, {\em provided} the BH is (a) reasonably bound to the cloud (initial velocity not larger than $\sim v_{\rm cl} \sim (G\,M_{\rm cl}/R_{\rm cl})^{1/2}$), and (b) the BH does not begin too close to the edge of the cloud with a velocity directed away from the (irregular) centers of collapse (in which case the BH tends to drift away from the dense regions, rather than interact with them). 

Another factor that could change the result is the initial metallicity $Z$, which self-consistently alters the cooling physics, stellar evolution tracks, and radiative feedback (opacities) in the simulations. We test this simulating GMCs with $M_{\rm cl}=10^6\,M_\odot$ and $R_{\rm cl}=5\,{\rm pc}$ (from the high surface density group) with varying initial $Z$ in Fig.~\ref{fig:metallicity}. By comparing the distribution functions of BH final-initial mass difference, we see that all those clouds produce statistically similar results for runaway BH accretion, independent of $Z$. We note that there are caveats regarding uncertainties in stellar evolution and treatment of molecular chemistry at extremely low metallicities in these models -- these are reviewed in detail in \citet{GrudicHopkins2018} -- but for all but truly metal-free clouds (where Pop-III evolution may be quite different) we regard this as robust. We also note that \citet{CorbettMoranGrudic2018} showed that the fragmentation and turbulent clumping in even metal-free clouds under high-density conditions like those of interest  here are quite similar, independent of different molecular chemistry networks used for the thermochemistry.

{The result does not mean the metallicity is not important for the actual accretion flow onto BHs, but is only valid for larger-scale accretion flows to the BH+disk system. Due to complexities of physics below our resolution limit ($r_{\rm sink}$), e.g., high metallicity may enhace the radiative force due to BH feedback and thus suppress accretion \cite[]{YajimaRicotti2017,ToyouchiHosokawa2019}.
}

We also change the number of BH seeds in the ICs ($N_{\rm bh,tot}$) and check the number of seeds that undergo significant mass growth, in Fig.~\ref{fig:bh-number-dependence}. Here we use different criteria to denote ``significant'': the final-initial mass ratio of the BH is above a constant $r$ and $r=100, 500, 2500$. If we simulate an initial number of seeds $N_{\rm bh,\,ini} \lesssim 64$, it becomes unlikely to see even a single seed undergo runaway growth, while for $N_{\rm bh,\,ini} \gtrsim 100$, we are essentially guaranteed that at least one seed will experience runaway growth. We find the same applying a more limited version of this test to other clouds. Thus there appears to be a threshold $\sim 1\%$ probability for a randomly-drawn seed to undergo runaway growth. However, if we increase the number of seeds further, the absolute number of BHs undergoing runaway accretion clearly saturates at a finite value, of $\sim$\,a few to ten with factor $10-100$ growth and $\sim 1$ with extreme runaway growth. Thus a given cloud can only support at most a few runaway BHs.

\section{Discussion}
\label{sec:discussions}

\subsection{Effects stellar feedback \&\ global cloud properties}
\label{sec:discussions:feeback}

\begin{figure*}
    \centering
    \includegraphics[width=\linewidth]{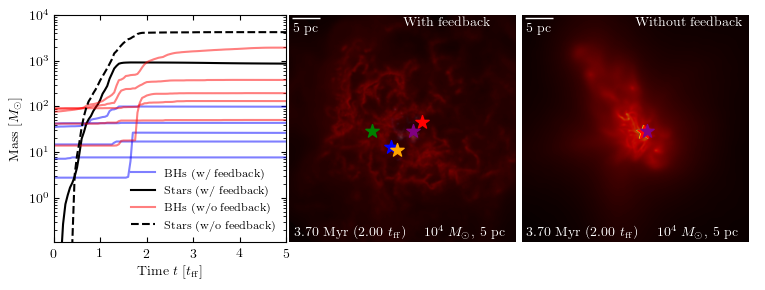}
    \vspace{-20 pt}
    \caption{Effects of stellar feedback on gas morphology and seed BH mass growth. We show two low-resolution runs of a low-density, low-mass cloud with $M_{\rm cl}=10^4~M_\odot$, $R_{\rm cl}=5~{\rm pc}$ cloud,  including (middle) and excluding (right) stellar feedback effects, as Fig.~\ref{fig:visualization}. Gas is more concentrated near the center for the no-feedback run while most gas is expelled by stellar winds in the feedback run. We show the cumulative BH and stellar mass growth for these runs as Fig.~\ref{fig:mass-growth} (left): BH growth and star formation are both suppressed at the order-of-magnitude level in these lower-density clouds.}
    \label{fig:feedback_effects}
\end{figure*}

Intuitively, stellar feedback can alter BH accretion in two ways. i) Feedback expels gas, which makes it harder for BHs to capture that gas. ii) Feedback can make the cloud more turbulent and create more dense regions. As an example, we include low-resolution simulations with and without feedback physics for the same ICs in Fig.~\ref{fig:feedback_effects}. As we see, for this low-surface density cloud, feedback effectively blows gas away after two free-fall time scales. As a result, feedback suppresses both BH accretion and star formation -- BH growth in particular is suppressed by more than an order of magnitude, the difference between there being a few versus essentially no ``runaway'' BHs.

Star formation and stellar feedback in GMCs have been well studied in previous simulations that are related to this work (e.g.\ \citealt{GrudicHopkins2018}), as well as similar studies with different numerical methods which have reached very similar conclusions for star formation (e.g.\ \citealt{2019MNRAS.487..364L}). One important conclusion of these studies (as well as more analytic ones like \cite{Larson2010}) is that the integrated star formation efficiency, and effects of feedback, depend strongly on $\bar{\Sigma}_{0}$. Briefly: a young stellar population of mass $M_{\ast}$ produces a net feedback momentum flux (from the combination of radiation and stellar winds) $\dot{P} \sim \langle \dot{p}/m_{\ast} \rangle\,M_{\rm cl,\,\ast} \sim 10^{-7}\,{\rm cm\,s^{-2}}\,M_{\rm cl,\,\ast}$, while the characteristic gravitational force of the cloud on its gas is $F_{\rm grav} \sim G\,M_{\rm cl,\,tot}\,M_{\rm cl,\,gas} / R_{\rm cl}^{2} \sim G\,M_{\rm cl,\,gas}\,\bar{\Sigma}_{0}$. So the gas reservoir of a cloud is rapidly unbound and ejected when its stellar mass exceeds $M_{\rm cl,\,ast}/M_{\rm cl,\,gas} \gtrsim G\,\bar{\Sigma}_{0} / \langle \dot{p}/m_{\ast} \rangle \sim \bar{\Sigma}_{0} / (1000\,M_{\odot}\,{\rm pc^{-2}})$. So for our low-$\bar{\Sigma}_{0}$ clouds, almost all of the gas is rapidly un-bound after just a small fraction of the GMC forms into stars, preventing it from being accreted by BHs.

\begin{figure*}
    \centering
    \includegraphics[width=\linewidth]{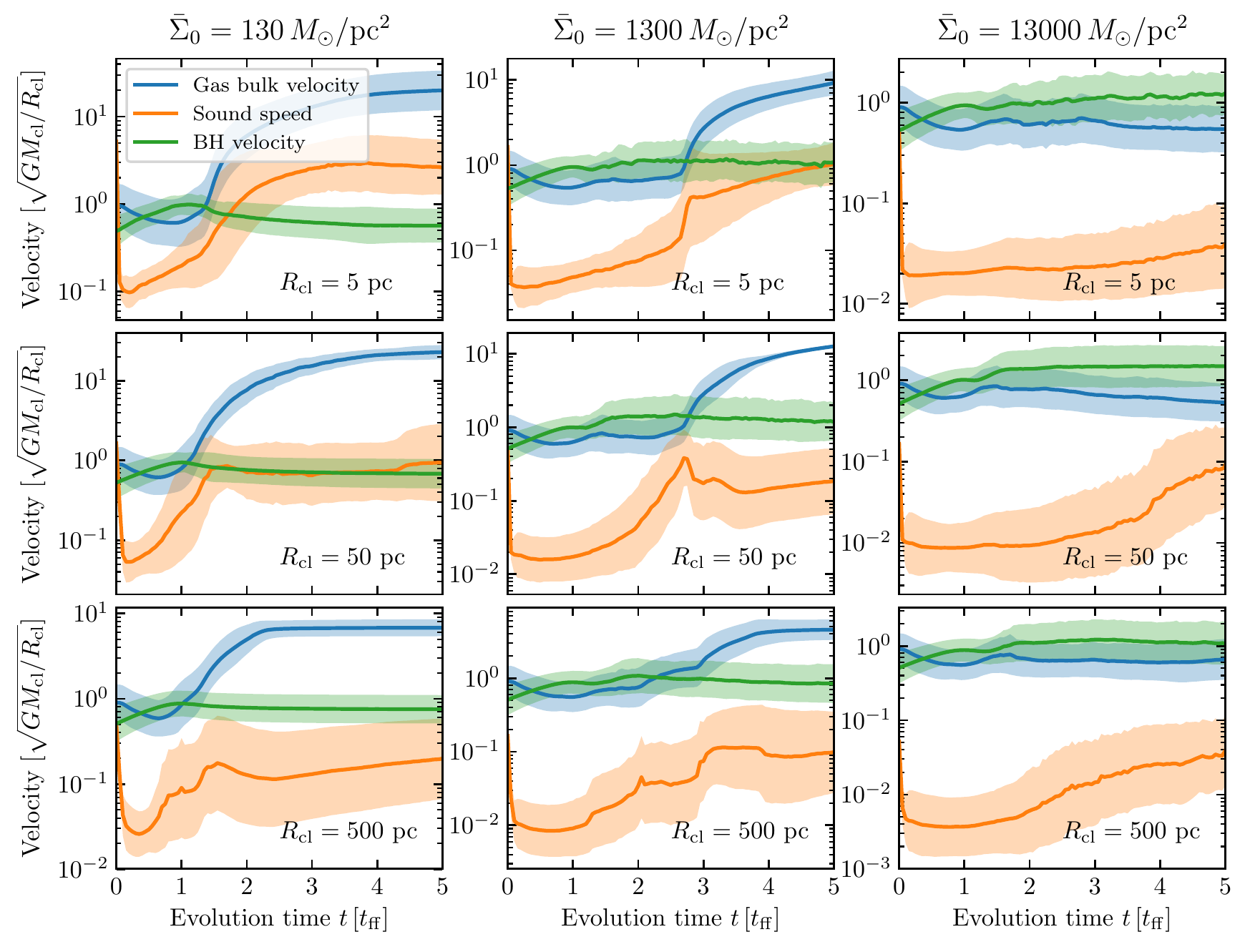}
    \vspace{-20 pt}
    \caption{Behavior of the rms gas bulk velocity dispersion ($\equiv  | \mathbf{v}_{\rm gas} - \langle \mathbf{v}_{\rm gas} \rangle| $), sound speeds ($c_{\rm s}$), and BH seed particle velocity dispersions relative to the complex center-of-mass ($\equiv  | \mathbf{v}_{\rm bh} - \langle \mathbf{v}_{\rm gas} |$), all in units of the characteristic gravitational/circular velocity of the complex ($v_{\rm circ} = v_{\rm cl} = \sqrt{G\,M_{\rm cl}/R_{\rm cl}}$), as a function of time (in units of the initial complex free-fall time ($t_{\rm ff}$). We compare the fiducial runs with different surface density (left-to-right) and radii (top-to-bottom). For each we show the median-absolute-value (line) and $14-86\%$ inclusion interval (shaded). Low-density, low-mass clouds are rapidly disrupted by stellar feedback producing large gas bulk velocities and high sound speeds. This is suppressed in high-density, high-mass systems, enabling continued BH accretion.}
    \label{fig:velocity-different-clouds}
\end{figure*}

We can see this reflected in Fig.~\ref{fig:velocity-different-clouds}, which shows the gas rms bulk velocity $\langle |\mathbf{v}_{\rm gas}|^{2} \rangle^{1/2}$, gas sound speed $\langle c_{\rm s} \rangle$, and BH rms velocity $\langle |\mathbf{v}_{\rm bh}|^{2}|\rangle^{1/2}$ as a function of time in different ICs, in units of the characteristic cloud gravitational velocity $v_{\rm cl} \sim (G\,M_{\rm cl}/R_{\rm cl})^{1/2}$. Not surprisingly, the rms velocity of BHs remains of order the gravitational velocities. The gas bulk velocities are dominated by gravity at first so remain of order $v_{\rm cl}$, but when feedback begins to disrupt the cloud they increase in magnitude by a factor $\sim 10$. This effect depends primarily on $\bar{\Sigma}_{0}$, as expected from the argument above.

Similarly, the sound speed of the clouds initially drops extremely quickly owing to cooling until it reaches molecular temperatures (we arbitrarily start from higher temperature, but this has no effect on our results), with $c_{\rm s} \ll v_{\rm cl}$, but then rises once feedback disrupts the cloud owing to a combination of (1) photo-ionization, (2) shocks from stellar winds bubbles, and (3) lower gas densities increasing the cooling time. Since e.g.\ the characteristic photo-ionized $c_{\rm s} \sim 10\,{\rm km\,s^{-1}}$ is roughly constant, the importance of this effect depends primarily on $v_{\rm cl}$, which ranges from $\sim 3\,{\rm km\,s^{-1}}$ in our lowest-mass, lowest-density simulation, to $\sim 300\,{\rm km\,s^{-1}}$ in our highest-mass, highest-density simulation.

In our low-density, low-mass clouds, we see disruption occurs very early (less than $\sim 2\,t_{\rm ff}$), with the gas bulk velocities and sound speeds reaching $\gg v_{\rm cl}$. This makes gravitational capture of gas by seeds nearly impossible. For the intermediate-density clouds, we see the disruption is significantly delayed, and the magnitude of the post-disruption velocities is reduced (with $c_{\rm s} \lesssim v_{\rm cl}$ even during disruption). For the highest-density clouds, there is no real disruption but just dispersal of some residual gas after star formation completes.

\begin{figure}
    \centering
    \includegraphics[width=\linewidth]{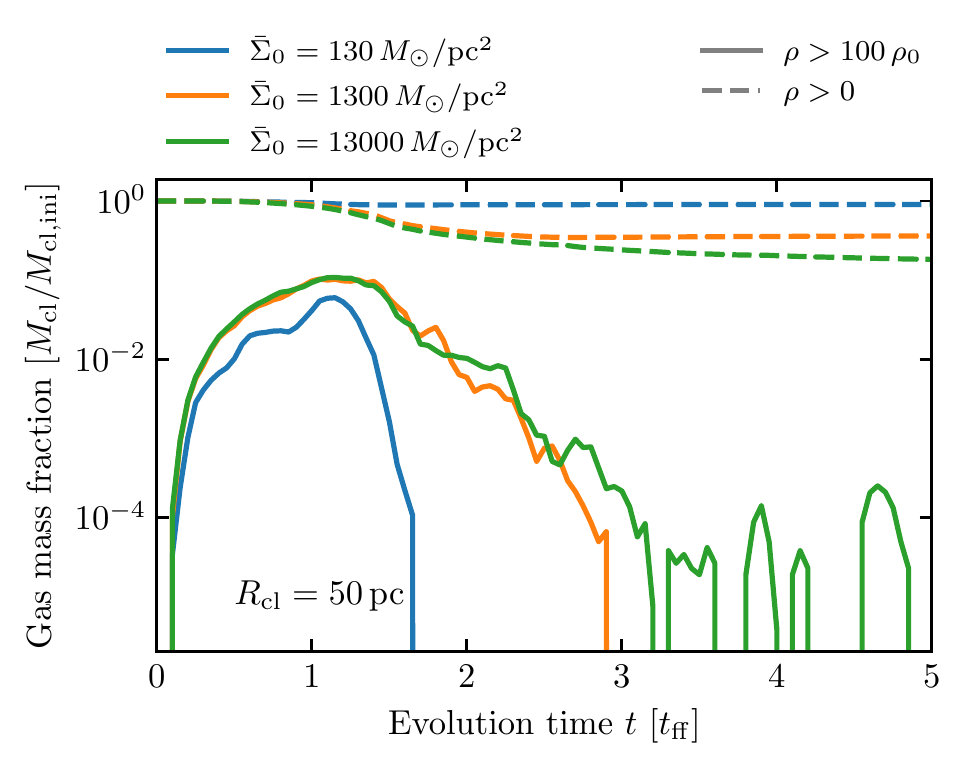}
    \vspace{-20 pt}
    \caption{Evolution of the abundance of dense clumps in the simulations. We compare complexes with different initial mean surface density $\Bar{\Sigma}_0$ (different colors; only $R_{\rm cl}=50\,{\rm pc}$ shown for simplicity). For each, we measure the total gas mass (in units of the initial mass $M_{\rm cl}$) with a local gas density exceeding $\rho > 100\,\langle \rho\rangle_{0}$ (where $\langle \rho \rangle_{0} \equiv 3\,M_{\rm cl}/4\pi\,R_{\rm cl}^{3}$ is the initial mean density). In all cases this rises rapidly as gravitational collapse and turbulence develop. In low-density systems, it almost immediately drops to zero after stellar feedback unbinds the gas. In higher-density systems, the presence of dense gas is sustained and even rejuvenated later in the system evolution owing to the presence of strong shocks.} 
    \label{fig:density-different-clouds}
\end{figure}

We have also considered the impact of stellar feedback on the volume and mass fraction in dense clumps (Fig.~\ref{fig:density-different-clouds}). Specifically, we calculated the volume and mass (in units of the initial cloud volume and total mass) of regions/clumps within the cloud that satisfy $\rho > 100\,\langle \rho \rangle_{0}$ (where $\langle \rho \rangle_{0} \equiv 3\,M_{\rm cl}/(4\pi\,R_{\rm cl}^{3})$ is the initial mean cloud density). The volume/mass of dense clumps increases in all cases rapidly at early times as the cloud collapses, but in the low-density clouds it is rapidly truncated by feedback. In contrast, we see in the higher-density clouds a sustained ``lifetime'' of dense gas: this is driven by shocks and turbulence from feedback from the stars that have formed, but have insufficient power to completely disrupt the cloud. In the highest-density case we even see dense clumps re-emerge several times after $t\gtrsim 3\,t_{\rm ff}$ due to large-scale stellar feedback events -- these correspond to large wind/HII region shells colliding to form dense regions (see e.g.\ \citealt{MaGrudic2020} for more detailed discussion). 

Another obvious requirement for runaway BH growth to ``interesting'' IMBH or even SMBH masses is that the total cloud mass is sufficiently large, such that the mass of dense ``clumps'' accreted is interesting. As we show below, the characteristic gas clump masses at high densities which meet the conditions for runaway BH growth are typically $\sim 1\%$ of the cloud mass, neatly explaining the maximum final BH masses seen in \S~\ref{sec:growth.vs.cloud}. This requires a total cloud mass $\gtrsim 10^{5}-10^{6}\,M_{\odot}$ for growth to true IMBH (let alone SMBH) scales. Interestingly, since $v_{\rm cl} \sim 15\,{\rm km\,s^{-1}}\,(M_{\rm cl}/10^{6}\,M_{\odot})^{1/4}\,(\bar{\Sigma}_{0}/10^{3}\,{\rm M_{\odot}\,pc^{-2}})^{1/4}$, this plus the surface density condition above simultaneously ensure that complexes are not over-heated or disrupted by photo-ionized gas.

\subsection{How Does Runaway Growth Occur?}
\label{sec:runaway}

We now consider the local conditions for runaway growth. In a small ``patch'' of cloud on scales $\sim r_{\rm sink}$ (small compared to the cloud but large compared to the BH accretion disk), it is not unreasonable to approximate the rate of gravitational capture of gas by a sink via the Bondi formula (Eq.~\eqref{equ:bondi-hoyle-rate}), given the local value of $\rho$, $\delta V$, and $c_{\rm s}$.

Recall, from \S~\ref{sec:intro} and \S~\ref{sec:simulations:bh-accretion}, that if we consider the typical or diffuse/volume-filling conditions within the cloud, i.e.\ $\rho \sim \langle \rho\rangle_{0} \sim 3\,M_{\rm cl}/4\pi\,R_{\rm cl}^{3}$ and $\delta V \sim v_{\rm cl} \gg c_{\rm s}$, we would obtain
\begin{align}
    \langle \dot{M}_{\rm bh} \rangle_{\rm diffuse}&  \sim \frac{G^2 M_{\rm bh}^2 \rho}{\delta V^3} \sim \frac{G^{1/2}\,M_{\rm bh}^{2}}{M_{\rm cl}^{1/2}\,R_{\rm cl}^{3/2}} \sim \left(\frac{\langle n_{\rm cl} \rangle}{{\rm cm^{-3}}}\right)\,\left( \frac{M_{\rm bh}}{M_{\rm cl}} \right)\,\dot{M}_{\rm Edd}
\end{align}
where $\langle n_{\rm cl}\rangle = \langle \rho \rangle_{0}/m_{p}$. If we further assume that the timescale for accretion $\Delta t$ is of order the cloud lifetime, $\sim t_{\rm eff} \sim \sqrt{R_{\rm cl}^3/GM_{\rm cl}}$, then the total mass accreted $\sim \dot{M}_{\rm bh}\,\Delta t$ would be
\begin{align}
\Delta M_{\rm bh} \sim \langle \dot{M}_{\rm bh} \rangle_{\rm diffuse}\,t_{\rm ff} \sim \frac{M_{\rm bh}}{M_{\rm cl}}\,M_{\rm bh}
\end{align}
In other words, unless the ``seed'' is already a large fraction of the entire GMC complex mass (i.e.\ is not really a seed in any sense), then the diffuse accretion will be highly sub-Eddington and the BH will grow only by a tiny fractional amount. This immediately explains why {\em most} of the seeds we simulate indeed grow negligibly.

However, in a highly turbulent cloud we argued above that two effects that may boost the mass growth: i) the dense clumps appear with $\rho \gg \langle \rho\rangle_{0}$, and ii) the turbulence velocity contributes to the relative velocity $\delta \mathbf{V}$ so locally, low $\delta V$ is possible.

\begin{figure}
    \centering
    \includegraphics[width=\linewidth]{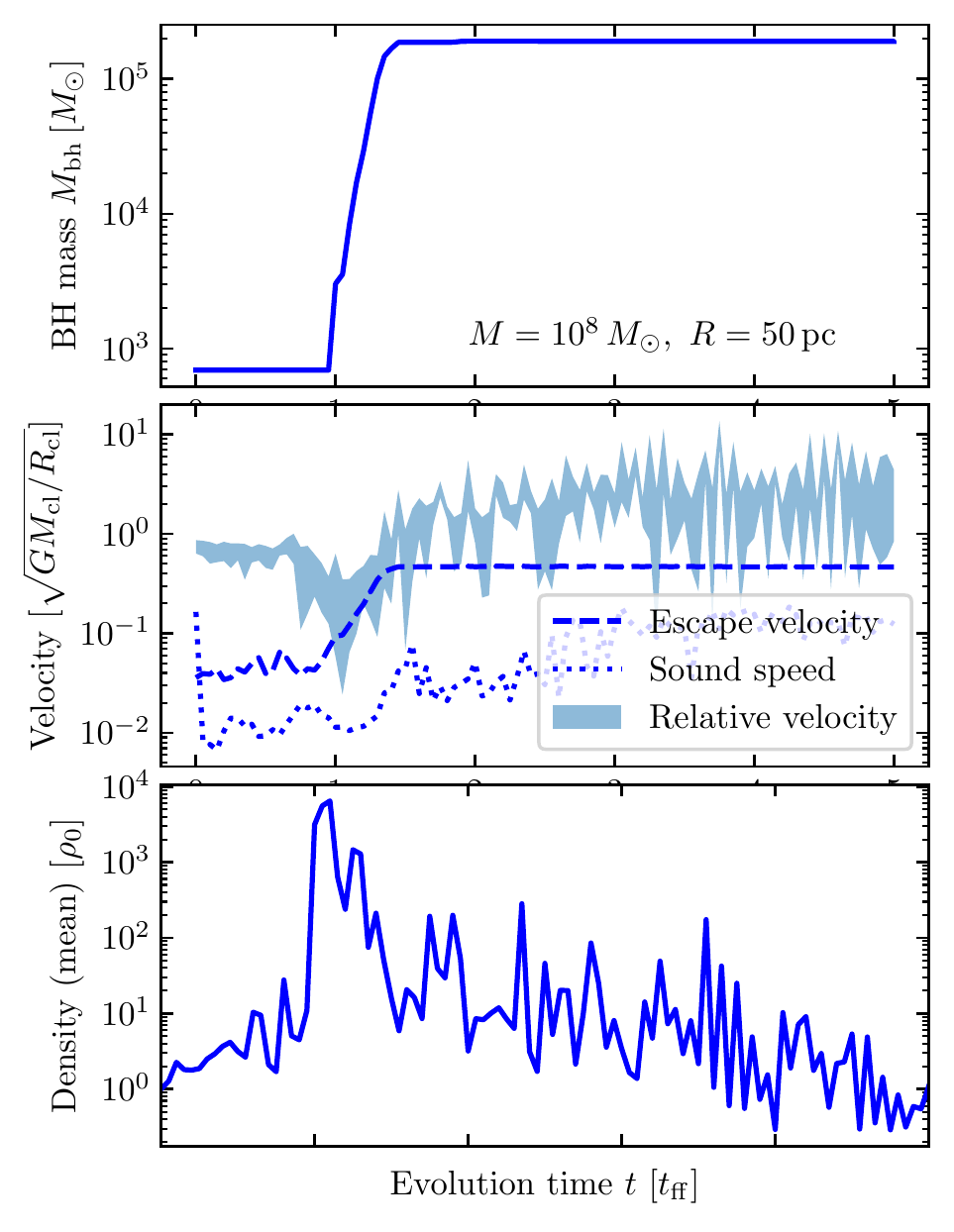}
    \vspace{-20 pt}
    \caption{A representative case study of the environment around one BH seed which undergoes runaway growth (complex properties labeled). 
    {\em Top:} BH/sink particle mass versus time (in units of the initial complex free-fall time $t_{\rm ff}$).
    {\em Middle:} Median sound speed ($c_{s}$; dotted), and $5-90\%$ range of relative gas bulk velocities ($|\mathbf{v}_{\rm gas}-\mathbf{v}_{\rm bh}|$; shaded) of all gas cells which fall within the sink radius $r_{\rm sink}$, and the escape velocity from the sink ($\sim \sqrt{G\,M_{\rm sink}/r_{\rm sink}}$). Velocities are in units of the characteristic complex gravitational velocity $v_{\rm cl}=\sqrt{G\,M_{\rm cl}/R_{\rm cl}}$.
    {\em Bottom:} Mean density of gas within the sink radius, in units of the initial complex mean density $\langle \rho \rangle_{0} \equiv 3\,M_{\rm cl}/4\pi\,R_{\rm cl}^{3}$. 
    The runaway growth occurs when the BH intercepts an overdense clump $\rho \gtrsim 100\,\langle \rho\rangle_{0}$, with a low relative velocity $\delta V \lesssim 0.1\,v_{\rm cl}$ so that it is gravitationally captured. Thermal pressure support/sound speed is relatively unimportant, since this is occurring in cold molecular gas.}
    \label{fig:bondi-hoyle-check}
\end{figure}

In Fig.~\ref{fig:bondi-hoyle-check} we follow one particular but representative example of a sink which undergoes runaway growth, considering how the relevant factors in the local Bondi rate evolve in the immediate vicinity of the sink. The thermal sound speed is negligible at basically all times in the cold molecular phases compared to $\delta V$, as expected. Runaway growth therefore occurs when the BH happens to encounter a region which simultaneously features a strong local density enhancement, $\rho \sim 10^{3}-10^{4}\,\langle \rho\rangle_{0}$, and low relative velocity $\delta V \lesssim 0.1\,v_{\rm cl}$, below the escape velocity of gas from the sink radius (so it is indeed captured). This boosts the local Bondi accretion rate by a factor of $\sim 10^{7}$, compared to our estimate above for the ``diffuse'' cloud medium. Visual examination shows this resembles Fig.~\ref{fig:visualization} -- the BH happens (essentially by chance) to be moving with a very low relative velocity to a dense clump created nearby by intersecting shock fronts (with Mach $\sim 30-100$ shocks, i.e.\ $v_{\rm shock} \sim 10\,{\rm km\,s^{-1}}$, producing the large density enhancement), and begins accreting it extremely rapidly. Since the Bondi rate scales as $\propto M_{\rm sink}^{2}$, this runs away and most of the clump mass ($\sim 10^{5}\,M_{\odot}$) is rapidly accreted and the clump is tidally disrupted and then captured before it fragments internally to form stars. 
Examination show this pattern is typical of the seeds which experience runaway accretion in the simulations.

\begin{figure}
    \centering
    \includegraphics[width=\linewidth]{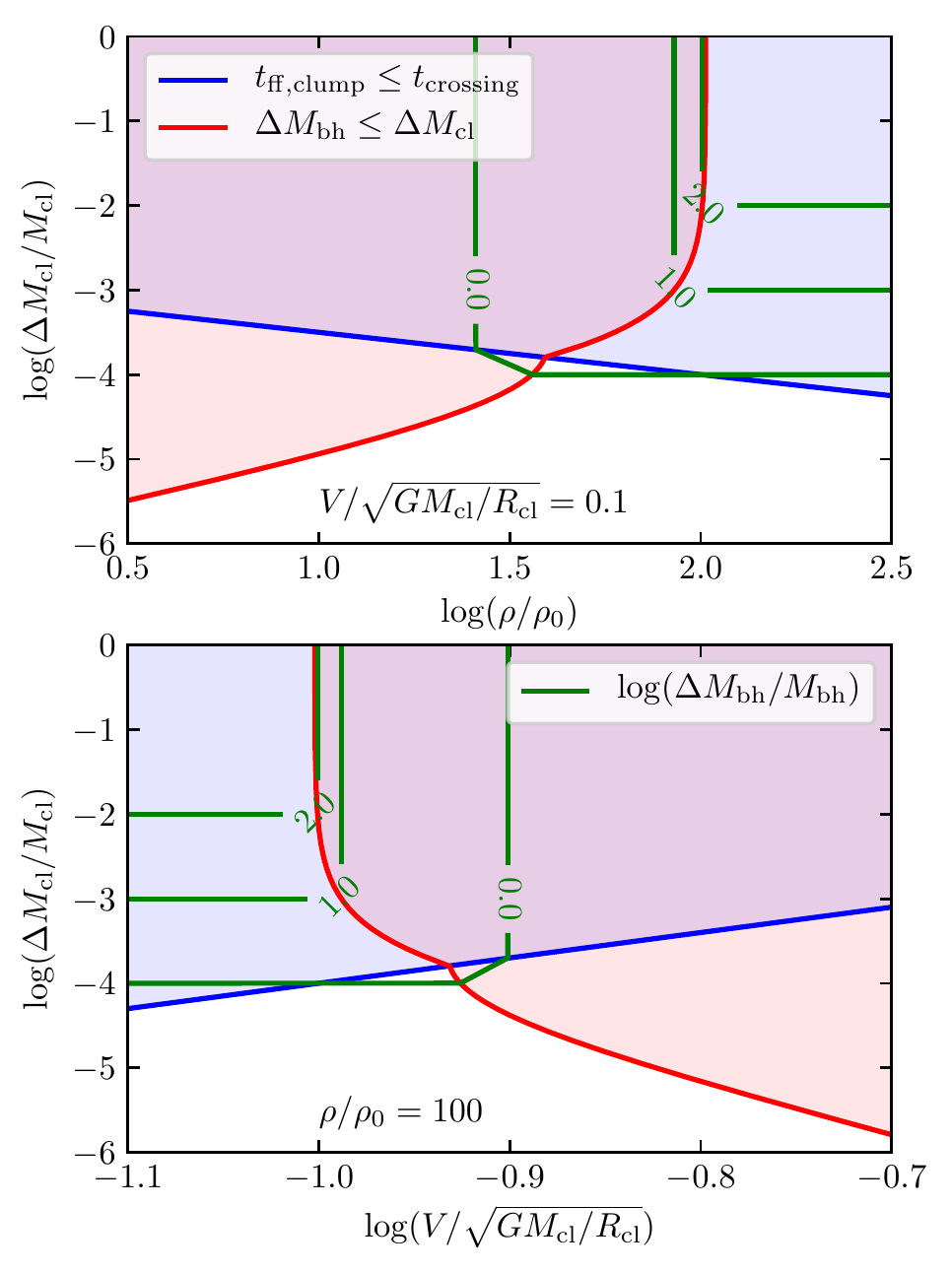}
    \vspace{-20 pt}
    \caption{Illustrative analytic predictions for BH accretion, based on a simple model of Bondi-Hoyle accretion in encounters with sub-clumps that have local density $\rho_{\rm c}$, mass $\delta M_{\rm c}$, and relative BH-clump velocity $\delta V_{\rm c}$. Here we assume the initial seed mass $M_{\rm bh} = 10^{-4}\,M_{\rm cl}$. 
    {\em Top:} Assuming a fixed $\delta V_{\rm c}$ (in units of $v_{\rm cl}$), we plot behavior as a function of $\delta M_{\rm c}$ and $\rho_{\rm c}$. 
    {\em Bottom:} Behavior as a function of $\delta M_{\rm c}$ and $\delta V_{\rm c}$, at fixed $\rho_{c}$. 
    The blue shaded region denotes where the internal free-fall time of the clump ($t_{\rm ff,\,clump}$) is shorter than the clump-BH crossing timescale ($t_{\rm crossing}$), so will be the rate-limiting timescale for accretion. 
    Red shaded range shows where the total mass accreted ($\Delta M_{\rm bh}$) over the shorter of $\Delta t = {\rm min}(t_{\rm ff,\,clump},\ t_{\rm crossing})$ would be less than the clump mass ($\delta M_{\rm c}$), so the accretion does not fully ``run away.''
    Green lines show contours where $\log_{10}(\Delta M_{\rm bh} / M_{\rm bh,\,0})$ is constant and equal to the value shown. 
    The region where $\Delta M_{\rm bh}/M_{\rm bh}$ contour is horizontal denotes ``runaway,'' defined by where the BH will accrete the entire clump mass.}
    \label{fig:analytic-scaling}
\end{figure}

Analytically, therefore, let us assume that during evolution, a BH encounters a dense clump with local density $\rho_{\rm c}$, clump radius $r_{\rm c}$, mass $\delta M_{\rm c}$, at relative velocity $\delta V_{\rm c}$ (and define $C^{2} = \delta V_{\rm c}^{2} + c_{\rm s}^{2}$, where we can generally assume $C \sim \delta V_{\rm c} \gtrsim c_{\rm s}$, even in regions where $\delta V_{\rm c}$ is relatively low), and accretes in Bondi-Hoyle-like fashion. Fig.~\ref{fig:analytic-scaling} summarizes the resulting accretion for various assumptions. Integrating the Bondi accretion rate for some time $\Delta t$ (assuming the background is constant), we have
\begin{align}
\label{equ:runaway-accretion}    \frac{1}{M_{\rm bh,0}} - \frac{1}{M_{\rm bh,final}} = \frac{1}{M_{\rm bh,0}} - \frac{1}{M_{\rm bh,0} + \Delta M_{\rm bh}}  & \sim \frac{4\pi\,G^2 \rho_{\rm c}}{C^{3}} \Delta t 
\end{align}
where $M_{\rm bh,\,0}$ is the ``initial'' BH mass. This diverges (formally $\Delta M_{\rm bh} \rightarrow \infty$), so in practice the entire clump will be accreted ($\Delta M_{\rm BH} \rightarrow M_{\rm c}$), in a finite time $\Delta t \rightarrow t_{\rm acc} \sim C^{3}/(4\pi\,G^{2}\,\rho_{\rm c}\,M_{\rm bh,0})$. In practice, the time $\Delta t$ will be limited by the shortest of either the dense clump lifetime (usually not shorter than its freefall time $t_{\rm ff,\,c} \sim 1/\sqrt{G\,\rho_{\rm c}}$), the timescale for the clump to fragment and form stars (also no shorter than $t_{\rm ff,\,c}$), or the crossing time $t_{\rm cross} \sim r_{\rm c} / \delta V_{\rm c}$ for the mutual interaction. A simple calculation shows that the ratio $t_{\rm cross}/t_{\rm ff,\,c} \sim (\delta M_{\rm c}/M_{\rm cl})^{1/3}\,(\rho_{\rm c}/\langle \rho\rangle_{0})^{1/6}\,(v_{\rm cl}/\delta V_{\rm c})$. Inserting numbers or considering Fig.~\ref{fig:analytic-scaling} shows that for the conditions of greatest interest for rapid accretion ($\delta V_{\rm c} \ll v_{\rm cl}$, $\rho_{\rm c} \gg \langle \rho\rangle_{0}$, and clump masses $\delta M_{\rm c}$ not incredibly small compared to the mass of the cloud so large BH growth is possible), we usually expect $t_{\rm cross} \gtrsim t_{\rm ff,\,c}$. So considering a ``worst-case'' scenario, then, accretion can run away to accrete the entire clump when $t_{\rm acc} \lesssim t_{\rm ff,\,c}$, which in turn requires: 
\begin{align}
\label{eqn:critical.condition} \frac{\delta V_{\rm c}}{v_{\rm cl}} \lesssim 0.1\,\left( \frac{M_{\rm bh,\,0}}{10^{-5}\,M_{\rm cl}} \right)^{1/3}\,\left( \frac{\rho_{\rm c}}{100\,\langle \rho \rangle_{0}} \right)^{1/6}
\end{align}
This corresponds reasonably well to the conditions where, in the simulations, we indeed see runaway growth -- regions with enhanced $\rho_{\rm c}$, and (crucially here), low local $\delta V_{\rm c}$. This also naturally explains why we see only a very weak dependence on initial BH mass -- provided this condition is met (which does not depend strongly on $M_{\rm bh,\,0}$), then the ``growth limiter'' is not the BH mass or Bondi rate (which depends strongly on $M_{\rm BH}$), but the mass of the clump $M_{\rm c}$ (which is, of course, entirely independent of the mass of the BH). Moreover, accretion events can occur sequentially, so once a BH ``jumps'' in mass by accreting a clump, its ``effective'' mass will be larger making it easier to accrete subsequent clumps (in an extreme form of competitive accretion). 

Still, if the BHs were truly extremely small (e.g.\ $\ll 10\,M_{\odot}$), or the clouds extremely massive, then the probability of such an event would become small rapidly -- this may explain, in part, why for the most massive complexes we see fewer BHs grow (but those that do grow, grow to even larger masses).

Finally, in Appendix~\ref{app:bh_accretion_rate_estimation}, we use this to make an order-of-magnitude estimate of the probability of a seed encountering a ``patch'' (i.e.\ clump) of gas meeting the criteria above. Assuming e.g.\ uncorrelated Gaussian velocity fields and lognormal density fields, we estimate that the probability of seeds encountering dense clumps is not  low, but the probability of such an encounter also having low relative velocity meeting the condition above is, giving a net probability in the range $\sim 0.001-0.01$. This is remarkably similar (given the simplicity of these assumptions) to our estimate of $\sim 0.01$ from the simulations where we varied the number of seeds systematically, as discussed above. 

{
\subsection{Hyper-Eddington accretion}

\begin{figure}
    \centering
    \includegraphics[width=\linewidth]{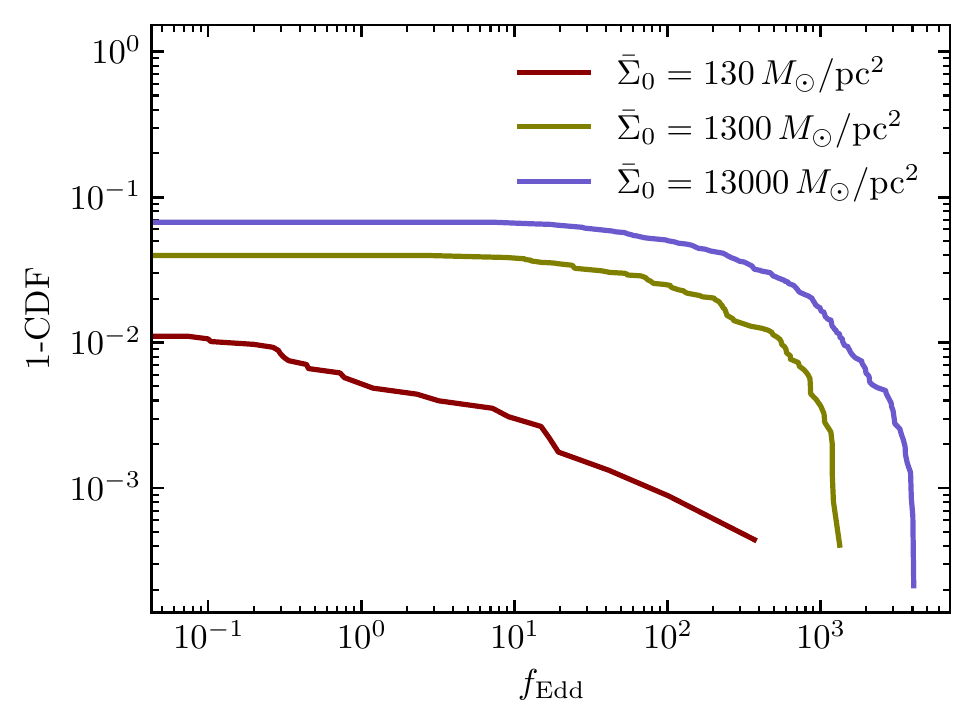}
    \vspace{-20pt}
    \caption{
    {Distribution of the Eddington ratio $f_{\rm Edd}$ for BHs in different groups of GMCs. Here for each BH, $f_{\rm Edd}$ is estimated in its fastest-accreting stage. For GMCs with high surface density, there are $\sim 2\%$ BHs reaching hyper-Eddington accretion with $f_{\rm Edd}\gtrsim 1000$. }
    }
    \label{fig:Edd-CDF}
\end{figure}

\begin{figure}
    \centering
    \includegraphics[width=\linewidth]{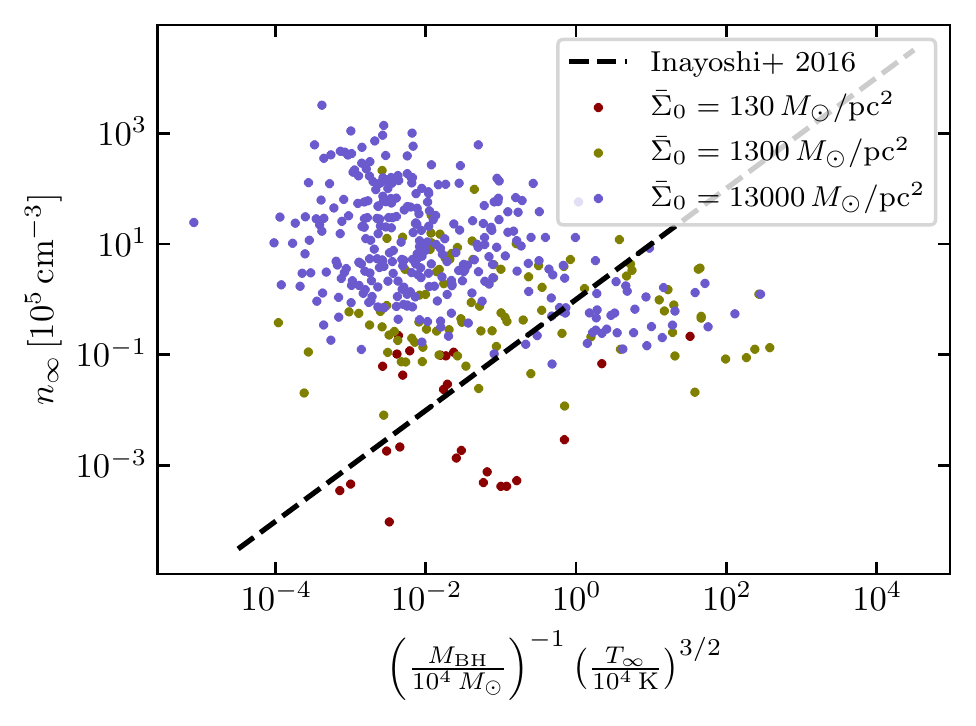}
    \vspace{-20pt}
    \caption{
    {Comparing the simulation with the critical density for hyper-Eddington accretion predicted by \citet{2016MNRAS.459.3738I}. Most BHs in the simulation, especially for those in GMCs with $\bar{\Sigma}_0=13000\,M_\odot/{\rm pc}^2$, are above the critical density. }
    }
    \label{fig:critical_density}
\end{figure}

Here we want to assess if the mass accretion onto BHs is hyper-Eddington. For this simulation without feedback, the mass flow onto BHs should already be enormous, which is a sufficient (though not necessary) condition for hyper-Eddington accretion. We first check this condition in Fig.~\ref{fig:Edd-CDF}. For each BH in the simulation, we can estimate its average mass accretion rate $\langle \dot{M}_{\rm BH} \rangle$ from neighboring snapshots. We define the Eddington ratio as $f_{\rm Edd} \equiv \langle \dot{M}_{\rm BH} \rangle / \dot{M}_{\rm Edd}$, and check the maximum $f_{\rm Edd}$ for each BH in its history. We then show the distribution of the maximum $f_{\rm Edd}$ for all BHs. There is a fraction of simulated BHs undergoing hyper-Eddington accretion (e.g., the fraction of BH with $f_{\rm Edd}\gtrsim 10^3$ is $\sim 2\%$ for GMCs with $\bar{\Sigma}_0 = 13000\,M_\odot/{\rm pc^2}$, and $\sim 0.5\%$ for those with $\bar{\Sigma}_0 = 1300\,M_\odot/{\rm pc^2}$). For BHs in GMCs with higher initial surface density, the possibility of hyper-Eddington accretion is also higher, in the same way as discussed in \S~\ref{sec:results}.

Feedback from black holes, especially the radiative feedback will play a negative role in mass accretion (details to be expanded in \S ~\ref{sec:caveats}). Although in this study BH feedback is not included, we can infer the availability of hyper-Eddington accretion from theoretical studies. \citet{2016MNRAS.459.3738I} predicted that the critical density for hyper-Eddington accretion (with the accretion rate of $\gtrsim 5000 L_{\rm Edd}/c^2$) is
\begin{align*}
    n_\infty \gtrsim 10^5 \left(\frac{M_{\rm BH}}{10^4\,M_\odot}\right)^{-1}\left(\frac{T_\infty}{10^4\,{\rm K}}\right)^{3/2} {\rm cm}^{-3}.
\end{align*}
Here $n_\infty$ and $T_\infty$ are the density and temperature near the BH. For each BH with mass accretion in our simulation, we track the time when the BH reaches the fastest accretion rate through its history, and measure the nearby gas density and temperature. We compare our simulation data with \citet{2016MNRAS.459.3738I} in Fig.~\ref{fig:critical_density}. We find that for most BHs the nearby density is above the critical density that allows hyper-Eddington accretion even if there is radiative feedback. For GMCs with higher surface density, the fraction of BHs above the critical density is also higher.

}

\subsection{{Effects of numerical parameters}}
\label{sec:discussions:hyper-Eddington}

At the end of the discussion section, we include the effects of several numerical parameters in the simulation.

\subsubsection{Mass \&\ Force Resolution}

Resolution could influence our results both directly and indirectly. Ideally we would wish to ensure $m_{\rm gas} \ll M_{\rm bh}$, and that the Bondi radii are resolved (see \S~\ref{sec:simulations:bh-accretion}), but there are of course physics we cannot hope to resolve in our larger cloud complexes, such as the formation of individual stars (e.g.\ predicting the IMF itself). Nonetheless, we have tested our results for several clouds at different resolution levels: $64^3$, $128^3$, and $256^{3}$ (See Appendix~\ref{app:resolution-convergence}). For most clouds (especially those with high initial mean surface density), we see no significant different in our statistical/qualitative conclusions across these resolution levels. Similarly, we have re-run two of our clouds (one low and one high density) with factor $\sim 3$ different force-softening values for the collisionless (star and BH) particles, and see no significant effect. Thus our conclusions appear robust, but we caution again that qualitatively new physical processes would need to be simulated if we went to higher resolution (which are represented here by our sub-grid models for e.g.\ IMF-averaged stellar evolution).

\subsubsection{BH Sink Radii}

As noted above, in the simulation the sink radius for BH accretion is set as the smaller value of the Bondi-Hoyle accretion radius with either $\delta V \sim c_{\rm s}$ or $\delta V \sim v_{\rm cl}$. Analysis of our simulation shows that runaway accretion almost always occurs in regions where $\delta V \sim c_{\rm s} \lesssim 0.1\,v_{\rm cl}$, with enhanced densities $\gtrsim 100\,(3\,M_{\rm cl}/4\pi\,R_{\rm cl}^{3})$, where (as noted above) the Bondi radii are orders-of-magnitude larger than one would calculate for the diffuse GMC gas with low relative velocities. As a result the simulations relatively easily resolve the Bondi radius where accretion is relevant. Nonetheless we have re-run a small subset varying $r_{\rm sink}$ by more than an order of magnitude in either direction. If we make $r_{\rm sink}$ much too small -- significantly smaller than the rms inter-cell separation (spatial resolution in the gas) at the highest cloud densities ($\gtrsim 100\,(3\,M_{\rm cl}/4\pi\,R_{\rm cl}^{3})$), then we artificially see suppressed accretion (simply because we require the separation between BH and gas cell center-of-mass be $<r_{\rm sink}$ for capture). If we make the sink radius more than an order of magnitude larger than our default value, we see spurious accretion, usually in the very early timesteps of the simulations (before turbulence is fully-developed), where diffuse gas with low relative velocities is rapidly accreted. But for more reasonable variations in $r_{\rm sink}$, even spanning an order of magnitude, our results are robust. And as we show below, the accretion corresponds fairly well with analytic post-processing expectations, further lending confidence to this estimate.

\subsubsection{Initial BH Velocities}

{We have considered some limited experiments where we add a systematic ``boost'' or arbitrarily increase the initial velocities of the BH seeds in the initial conditions (for details see Appendix~\ref{app:initial_velocity_dependece})}. As expected, if the seeds are moving well in excess of the escape velocity relative to the cloud, they escape rapidly and never capture a large amount of gas. So the rms velocities of ``interesting'' BH seeds can only exceed $v_{\rm cl}$ by a modest (at most order-one) factor. On the other hand, reducing the BH seed velocities to zero has very little effect (other than introducing some spurious transient accretion in the first few timesteps when there is no relative gas-BH motion), because they quickly acquire random velocities of order the gravitational velocity $v_{\rm cl}$ from the fragmenting cloud potential.

\subsection{{Connections with observations}}

Given that this is really a theoretical ``proof of concept'' and we do not yet include these crucial physics (which we expect may change the key conclusions), we hesitate to make specific observational predictions. Nonetheless, even if BH feedback did nothing to further suppress runaway BH growth, there are some important conclusions we can draw regarding observations of both active (star-forming) clouds and ``relic'' star clusters. 
\begin{enumerate}
\item Runaway accretion would not occur in Milky Way/Local Group GMCs or cloud complexes: the necessary conditions much more closely resemble starburst galaxy nuclei and the most massive dense star-forming clumps observed in high-redshift massive galaxies.
\item As a result, the ``relic'' star clusters from regions which could produce runaway accretion will not be typical star clusters or globular clusters. They are much more akin to nuclear star clusters (at the low-mass end) and dense galactic bulges (at the high-mass end). Even if the high-redshift clumps are off-center, these complexes would quickly spiral to the center of the galaxies to form proto-bulges \citep{noguchi:1999.clumpy.disk.bulge.formation,dekel:2009.clumpy.disk.evolution.toymodel}, which is important for SMBH seed formation mechanisms as it is almost impossible for seeds of the masses we predict here to ``sink'' to galaxy centers via dynamical friction in a Hubble time at high redshift, if they are not ``carried'' by more massive star cluster complexes \citep{ma:2021.seed.sink.inefficient.fire}.
\item Regardless of which clusters could have hosted this runaway process, we again find the probability is low on a ``per seed'' basis. Therefore, whether we expect an IMBH/SMBH ``relic'' in the descendants depends fundamentally on the population of seeds and their dynamics. While we find stellar-mass seeds are viable, it is not obvious if these could come from the stars forming in the cloud itself (e.g. from the relics of the stars formed during the process). Most stellar-mass seeds form relatively late after star formation ($\gtrsim 30\,$Myr), in explosions (which could disrupt the cloud), and have large natal kicks (excessive relative velocity). It is possible, if kicks were somehow avoided, that the most massive stars which reach the end of the main sequence more rapidly (at $\sim 3\,$Myr) and collapse directly to BHs could be viable seeds, but then these are much more rare. Alternatively, the seeds could come from the ``pre-existing'' background of stars, as especially in e.g.\ galactic nuclei or $\sim$\,kpc-scale clump complexes in massive galaxies we expect a very large population of background stellar-mass BHs. The key then is their kinematics (i.e.\ whether a sufficient number can be ``captured'' to locally interact as we model).
\item Almost by definition, the required conditions make it very difficult to observe this process ``in action.'' Complexes which meet the criteria above are, by definition, Compton-thick (and since the accretion occurs in over-dense sub-regions, these are even more heavily obscured). Moreover, if the maximum luminosity of accreting BHs (even if they are undergoing hyper-Eddington accretion) is not much larger than the traditional Eddington limit (as most models predict; see \S~\ref{sec:intro}), then the bolometric and even X-ray luminosities of the clouds/complexes will be dominated by the stars (not the runaway accreting BHs), unless the BH accretes an enormous fraction ($\sim 10\%$) of the entire cloud mass. 
\item Even if such enormous accretion were to occur (or if the luminosity could exceed Eddington), by the time the BH luminosity could ``outshine'' even a fraction of the stellar luminosity of the complex, its luminosity would be so large that it would not be a ULX-type source. Rather (especially, again, recalling that the complexes of interest are generally in or around distant galactic nuclei), it would much more closely resemble an off-center, obscured AGN (or a dual AGN, if the galaxy already has an accreting SMBH). Large populations of such AGN sources are, of course, well-known, and there are much more mundane ways to produce them (via galaxy mergers or irregular kinematics), but it is perhaps conceivable that a small fraction of them could be systems like what we simulate here.
\end{enumerate}

\section{Caveats}
\label{sec:caveats}

\subsection{Feedback from Accreting Black Holes}

The most important caveat of this study is that we did not include any ``sub-grid'' model for BH accretion or feedback in the simulations. So ``BH accretion rate'' here should really be understood to be ``rate of gravitational capture of gas by the BH-disk system'' (akin to ``Bondi-Hoyle-like mass inflow rate'') and ``BH mass'' or ``sink mass'' represents a sum of the actual BH mass and its bound/captured material (whether that material has actually formed an accretion disk is another question itself).

This is not, of course, because we expect feedback to be unimportant for the BHs which rapidly capture gas: indeed, models of super-Eddington accretion disks (models whose ``outer boundary condition'' is something like the sink radii or ``inner boundary condition'' of our simulations) predict both strong radiative (luminosities near or somewhat above the Eddington luminosity) and kinetic (broad-angle MHD outflows from the disk) feedback (see references in \S~\ref{sec:intro}). While it is conceivable that under sufficiently-dense conditions, the surrounding material could continue to accrete (see e.g.\ \citealt{QuataertGruzinov2000,TakeoInayoshi2018,ReganDownes2019}), this could also completely shut down BH growth and even star formation in the surrounding cloud \citep{SchawinskiKhochfar2006}. 

However, crucial details of such accretion and feedback processes remain deeply uncertain. This includes (i) the rate at which material can go from being ``gravitationally captured'' to actually accreted onto the BH (which determines the luminosity and other feedback); (ii) whether star formation and/or fragmentation occurs in the captured disk material if too much mass is captured; and (iii) for a given accretion rate, the radiated spectrum and energy, and the energy and momentum and mass and opening angle of accretion-disk winds. Our intention here is therefore to first identify a set of {\em necessary}, but perhaps not sufficient, pre-conditions for runaway hyper-Eddington seed growth in on ISM scales. Clearly, if a BH cannot sustain super-Eddington gravitational capture rates of sufficient total mass in the first place, then it is unlikely that adding feedback would do anything other than further decrease the (already minimal) accretion. This allows us to already rule out large segments of parameter space as viable locations for hyper-Eddington accretion (e.g.\ Milky Way-like low-density or low-mass clouds, systems with insufficient statistical sampling of ``seeds,'' highly-unbound seeds). In future work (in preparation), we will use this study as the basis for a follow-up set of simulations which do include BH feedback, systematically surveying parameterized models for the major uncertainties described above, but using the results of this study to specifically begin from conditions where we know that, {\em absent} BH feedback, rapid accretion is possible.

\subsection{Other Caveats}

There are also of course other caveats in our simulations themselves. While we survey a factor of $\sim 100$ in mass resolution and see robust results, we are certainly not able to ``converge'' in a strict sense, especially given some ISM micro-physics where the relevant dynamics occur on sub-au scales. We cannot resolve or predict the IMF or stellar evolution tracks of individual stars, let alone their late-stage evolution and potential collapse into BH relics. This is especially unfortunate as one might imagine one source of ``seed'' BHs bound to the cloud would be extremely massive stars that form in that cloud with very short lifetimes that might implode or collapse directly to massive BH remnants, rather than exploding as SNe. A new generation of simulations like STARFORGE might be able to address some of these, but the resolution required has thus far limited their explicit simulations to precisely the low-density, low-mass clouds of least interest here \citep{GrudicGuszejnov2021,GrudicKruijssen2021}. 

It is also possible that physics we neglect plays an important role. For example, on galactic scales, cosmic rays can influence the ISM significantly, although many have argued that because of their diffusive nature (smooth CR density gradients), they play little dynamical role (except perhaps via ionization) in the dense ISM clouds of interest here \citep{farber:decoupled.crs.in.neutral.gas,hopkins:cr.mhd.fire2,hopkins:cr.multibin.mw.comparison,bustard:2020.crs.multiphase.ism.accel.confinement}. 

More realistic initial conditions and boundary conditions for clouds (embedded in a full ISM, for example) could also be important \citep{lane:2022.turbcloud}. This is perhaps especially relevant for our most massive complexes. When we simulate regions with $R_{\rm cl} \sim 500\,$pc and $\bar{\Sigma}_{0} \sim 10^{4}\,{\rm M_{\odot}\, pc^{-2}}$ -- i.e.\ gas masses as large as $\sim 10^{10}\,M_{\odot}$, these are not really ``clouds'' as we think of GMCs in the Milky Way. Rather, these values are motivated by typical sizes and densities observed in systems like starburst and/or ultra-luminous infrared galaxy nuclei \citep[see e.g.][]{kennicutt98,gao:2004.hcn.compilation,narayanan:2008.sfr.densegas.corr,bigiel:2008.mol.kennicutt.on.sub.kpc.scales}, and seen in the common massive clump-complexes or nuclei of high-redshift massive galaxies \citep{tacconi:high.molecular.gf.highz,krumholz:2010.highz.clump.survival,narayanan:2011.z2.kslaw,orr:ks.law}. But under these conditions, there is usually also a large pre-existing stellar population and dark matter halo, defining the potential of the nuclear gas -- properly simulating these regimes would really require full galaxy-formation simulations. It is likely that this added potential would make the starburst even less able to disrupt, leaving behind a dense nuclear bulge \citep[e.g.][]{sanders88:quasars,tacconi:ulirgs.sb.profiles,rj:profiles,hopkins:cusps.mergers,hopkins:cusps.fp,hopkins:cusps.evol,hopkins:cusps.ell,hopkins:cusp.slopes}.

\section{Conclusions}
\label{sec:conclusions}

We have simulated populations of dynamic, accreting BH seeds with masses $\sim 10^{1}-10^{4}\,M_{\odot}$ in massive cloud complexes (meant to resemble the most massive GMCs, high-redshift and starburst galaxy nuclei), with self-gravity, realistic cooling, detailed models for star formation and stellar feedback in the form of radiation, outflows, and supernovae, but neglecting the feedback from the BHs themselves. Our goal is to identify whether, under any conditions, such seeds can capture gas from the dense, cold ISM at the rates required to power hyper-Eddington accretion, and whether this can be sustained for long enough periods that it is conceivable such BHs could grow to IMBH or even SMBH mass scales. This forms a necessary, but not sufficient, condition for hyper-Eddington growth models for the origin of IMBHs and SMBHs. 

Based on our analysis above, we can draw the following conclusions (again, absent BH feedback): 
\begin{enumerate}
\item Sustained hyper-Eddington gravitational capture from the ISM can occur, under specific conditions (detailed below). This occurs when BH seeds coincidentally find themselves in regions with locally enhanced densities (local densities well in excess of $\sim 100$ times the complex-mean), with (by chance) very low relative velocities (less than $\sim 10\%$ of the characteristic gravitational velocity of the complex). The dense clump is then captured extremely quickly (on less than its internal dynamical time), which can set of a ``runaway'' of competitive accretion by which the seed grows even more massive (reaching up to $\sim 1\%$ of the complex gas mass).
\item Provided the right conditions are met, this process is only very weakly dependent on the initial seed mass, even for stellar-mass seeds in the $\sim 10-100\,M_{\odot}$ range. Thus, the ``seed'' does not need to already be an IMBH. 
\item Much like with star formation, stellar feedback plays a dual role. Stellar feedback overall suppresses star formation and unbinds gas, suppressing BH growth (especially in lower-density clouds). But in higher-density, more-massive complexes, feedback produces regions like colliding shocks/shells which promote exactly the conditions needed for runaway BH growth.
\end{enumerate}

For this runaway accretion to occur, we show that there are several necessary ``global'' criteria the molecular complex must meet, including: 
\begin{enumerate}
\item The complex must have a high surface density/gravitational pressure, $\bar{\Sigma}_{0} \gtrsim 1000\,{\rm M_{\odot}\,pc^{-2}}$. Otherwise, stellar feedback disrupts the medium too efficiently, both reducing the time available for accretion but also unbinding dense gas instead of allowing it to remain trapped and thus potentially creating situations with low relative velocities. 
\item The complex must also be sufficiently high-mass, $M_{\rm cl} \gtrsim 10^{6}\,M_{\odot}$. This is to ensure both that there is sufficient total mass supply that if hyper-Eddington accretion occurs, the final mass is ``interesting'' (reaching IMBH, let alone SMBH mass scales), but also required, along with the  $\bar{\Sigma}_{0}$ criterion, to ensure that the escape velocity of the cloud will be large enough that ionizing radiation does not rapidly unbind material or disrupt the complex and prevent accretion. 
\item The BH seeds must be ``trapped'' by the complex, with systematic relative velocities not significantly larger than the characteristic gravitational velocity of the cloud. This means, for example, that a BH moving isotropically in the background galaxy bulge, intersecting a cloud, would be unable to accrete, while BHs with small relative velocities to the cloud are viable.
\item We require at least $\sim 100$ seeds, in complexes meeting all the criteria above, to have an order-unity probability of one showing sustained hyper-Eddington accretion. Thus even when all the criteria are met, the conditions are ``rare'' on a per-seed basis. Once the number of seeds is sufficiently large, the finite number of locations where runaway can occur, plus the competitive accretion dynamics noted above, mean that the number which actually do experience runaway growth saturates at one to a few.
\end{enumerate}

In future work, we will use this preliminary study to inform a more focused study which does include BH feedback, systematically exploring the uncertainties in BH feedback models but focused on cloud conditions where -- at least in the absence of said feedback -- we find runaway growth is possible.

\acknowledgments{We thank Xiangcheng Ma and Linhao Ma for useful discussions and revisions of this draft. Support for the authors was provided by NSF Research Grants 1911233, 20009234, 2108318, NSF CAREER grant 1455342, NASA grants 80NSSC18K0562, HST-AR-15800. Numerical calculations were run on the Caltech compute cluster ``Wheeler,'' allocations AST21010 and AST20016 supported by the NSF and TACC, and NASA HEC SMD-16-7592.}

\datastatement{The data supporting the plots within this article are available on reasonable request to the corresponding author. A public version of the GIZMO code is available at \gizmourl.}

\bibliography{bib}

\begin{appendix}

\section{Estimating the Probability of a Runaway Accretion Event}
\label{app:bh_accretion_rate_estimation}

Based on arguments in \ref{sec:discussions}, we try here to estimate the probability of a runaway BH accretion event. Specifically, from \S~\ref{sec:runaway}, we want to estimate the probability of a random BH seed encountering a ``clump'' in a turbulent cloud complex which meets the conditions defined in Eq.~\ref{eqn:critical.condition}. For simplicity (although somewhat motivated by simulations, see e.g.\ \citet{burkhart:2009.mhd.turb.density.stats}) we will assume uncorrelated density and velocity fields, with Gaussian velocity statistics and lognormal density statistics as is usually assumed in supersonic turbulence. 

First consider the probability of a seed encountering a clump which is overdense in the manner of Figs.~\ref{fig:visualization} \&\ \ref{fig:bondi-hoyle-check}, or Eq.~\ref{eqn:critical.condition}. Assume the seeds have an rms velocity dispersion $\langle \mathbf{v}_{\rm bh}^{2} \rangle^{1/2}$ of order the gravitational velocity of the complex $v_{\rm cl}$, as does the gas, and that the complex is filled with some number density of clumps $n_{\rm c}$ with effective cross-section $\sigma_{\rm c} \sim {\rm Vol}_{\rm c} / r_{\rm c}$, where ${\rm Vol}_{\rm c}$ is the volume of a typical clump. Assume the seeds randomly move through the complex (uniformly sampling the volume) over its lifetime $\Delta t = \tau\,t_{\rm ff,\,cl}$ (where for the massive complexes of interest, $\tau \sim$\,a few, and $t_{\rm ff,\,cl} \equiv R_{\rm cl}/v_{\rm cl}$). The average number of dense clumps encountered is therefore $\langle N_{\rm cl} \rangle \sim \tau\,n_{\rm c}\,\sigma_{\rm c}\,v_{\rm cl}\,t_{\rm ff,\,cl} \sim \tau\,(R_{\rm cl}/r_{\rm c})\,f_{\rm V,\,c}$. We are interested in dense clumps or shocks, which simulations show tend to have a characteristic size/width of order the sonic scale, $r_{\rm c} \sim R_{\rm cl}/\mathcal{M}_{\rm comp}^{2}$ (where $\mathcal{M}_{\rm comp}$ is the compressive Mach number of the cloud; see e.g.\ \citealt{passot:1988.proof.lognormal,Vazquez-Semadeni1994,scalo:1998.turb.density.pdf}). We can also estimate $f_{\rm V,\,c}$ by integrating the standard volume-weighted lognormal density PDF for supersonic turbulence (Gaussian in $\ln{(\rho / \langle \rho\rangle_{0})}$, with mean $=-S/2$ required by continuity, and variance $S \approx \ln{[1 + \mathcal{M}_{\rm comp}^{2}]}$) above some critical $\rho \gtrsim \rho_{\rm c} \sim 100\,\langle \rho \rangle_{0}$. Plugging in numbers, we can see that $\langle N_{\rm cl} \rangle \gtrsim 1$ so long as $\mathcal{M}_{\rm comp} \gtrsim 10$, which is easily satisfied in cold molecular gas for the massive, high-density complexes of interest.

So it is not particularly rare for a BH to encounter a dense clump over a duration of several free-fall times in the massive, dense complexes of interest. What is less common is for such an encounter to feature a low relative velocity $\delta V = |\mathbf{v}_{\rm bh} - \mathbf{v}_{\rm gas}|$. Let us assume, similar to the above, that the BH velocity is drawn from an isotropic Gaussian distribution with 1D dispersion $\sigma_{\rm v,\,bh} \sim v_{\rm cl}/\sqrt{3}$ in each dimension, and the gas or clump velocity is drawn from an independent isotropic Gaussian with similar 1D dispersion in each dimension $\sigma_{\rm v,\,gas} = \alpha_{v}\,\sigma_{\rm v,\,bh}$ (where $\alpha_{v}$ is an arbitrary order-unity constant). The velocity difference $\delta \mathbf{V}$ is therefore also Gaussian-distributed, and integrating we obtain the probability
\begin{align}
P_{v}(\delta V < \epsilon\,v_{\rm cl}) &= {\rm erf}\left(\frac{q}{\sqrt{2}}\right)-\sqrt{\frac{2}{\pi}}\,q\, \exp\left(-\frac{q^2}{2}\right)\ , \\ 
q &\equiv \frac{\sqrt{3}\,\epsilon}{\sqrt{1+\alpha_{v}^{2}}}
\end{align}
Assuming $\alpha_{v} \sim 1$, and $\epsilon \sim 0.1$ from Eq.~\ref{eqn:critical.condition}, we obtain $P_{v} \sim 5\times10^{-4}$; multiplying by $\langle N_{\rm cl} \rangle \sim$\,a few (for our massive cloud complexes), this gives a probability of $\sim 10^{-3} - 10^{-2}$ of an ``interesting'' event, per seed. 

We stress that this is only intended as a guide for intuition -- we have ignored a wide range of effects which modify the statistics, the fact that the density and velocity statistics are probably correlated \citep[see e.g.][]{konstandin:2012.lagrangian.structfn.turb,squire.hopkins:turb.density.pdf}, the fact that strong shocks and feedback tend to produce large local deviations from Gaussianity \citep{hopkins:2012.intermittent.turb.density.pdfs,beattie:2021.turb.intermittency.mhd.subalfvenic}, gravitational focusing (which probably significantly increases the rate of ``coincidences'' in velocity-density space), the size spectrum of different density structures \citep{guszejnov:gmc.to.protostar.semi.analytic,guszejnov:universal.scalings}, and more. Ultimately, capturing all of these effects is only possible in the full simulations, but the simple arguments here can provide some very approximate guide to the typical behaviors in the simulations.

\section{Resolution convergence}
\label{app:resolution-convergence}

\begin{figure}
    \centering
    \includegraphics[width=\linewidth]{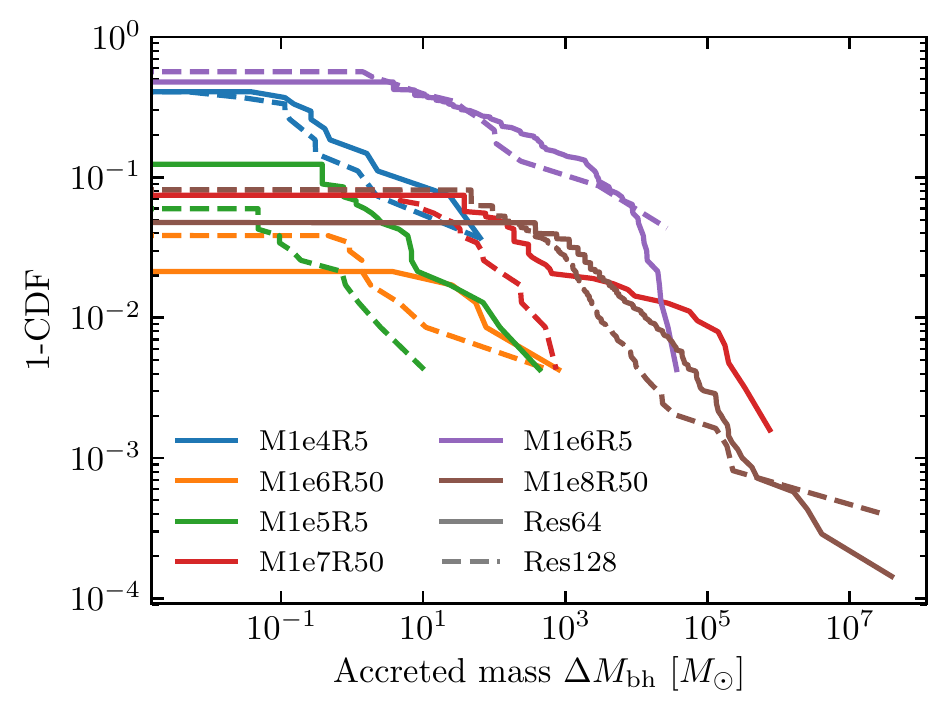}
    \vspace{-20 pt}
    \caption{Cumulative distribution of accreted mass as a resolution convergence test. Here we show clouds with initial radius $R_{\rm cl} \le 50\,{\rm pc}$. Each cloud is simulated with resolutions of $64^3$ (solid) and $128^3$ (dashed). We see general good agreement for simulations in different resolution for most clouds.
    }
    \label{fig:resolution-test}
\end{figure}

In Fig.~\ref{fig:resolution-test} we show the cumulative distribution of BH final-to-initial mass difference (same as Fig.~\ref{fig:cdf}) for simulations in different resolutions. Here we choose clouds with $R_{\rm cl}\le 50\,{\rm pc}$, such that all clouds weigh $M_{\rm cl} \le 10^8\,M_\odot$ and the condition $m_{\rm gas} \ll M_{\rm bh}$ is more likely to be satisfied. We see that for low (\texttt{M1e4R5}\footnote{In this section, we use the template \texttt{M\%eR\%d} to denote a cloud with its mass (in $M_\odot$) and radius (in pc). } and \texttt{M1e6R50}) and high (\texttt{M1e6R5} and \texttt{M1e8R50}) surface density clouds, there is good agreement in the CDFs under high (\texttt{Res128}) and low (\texttt{Res64}) resolutions. The resolution convergence is worse for the medium surface density group: for \texttt{M1e7R50} we see the same cut-off but different span of CDFs, while for \texttt{M1e7R50} there is a systematic difference. One possible reason might be the uncertainties in these near-breakup clouds. 

As a summary for the test, we find good quantitative resolution convergence for most clouds, especially dense ones where there is significant BH accretion (\texttt{M1e6R5} and \texttt{M1e8R50}). There are indeed some clouds with systematic offsets under different resolutions, but they fall into a ``less-interesting" category in terms of BH accretion and will not qualitatively change our conclusions.

{
\section{Initial velocity dependence}
\label{app:initial_velocity_dependece}

Due to considerations in computational costs, we made some confinements in the initial velocity distribution in order to capture BH accretion events more efficiently in a limited suite of simulations. In the main text the fiducial choice is to have all BH initial velocity $v_{\rm ini}$ below the circular velocity $v_{\rm cl}\equiv (GM_{\rm cl}/R_{\rm cl})^{1/2}$. Inevitably this will miss some BHs that are still bounded to the GMC, and possibly overestimate (or underestimate) the possibility of runaway accretion in a more general BH seed population. In this section we inspect the issue with a test. 

In Fig.~\ref{fig:new_velocity} we show the initial-velocity depends and cumulative distributions of BH accretion (featured by the mass accreted by each BH, $\Delta M_{\rm BH}$), for three cutoffs in the initial velocity magnitude: fiducial ($v_{\rm ini} \le v_{\rm cl}$), ``critically bounded'' ($v_{\rm ini} \le \sqrt{2} v_{\rm cl}$, which means some BHs may reach $v_{\rm ini} \lesssim \sqrt{2} v_{\rm cl}$ ), and ``unbounded'' ($v_{\rm ini} \le \sqrt{2} v_{\rm cl}$). For each test, other quantities like the initial position and velocity direction for each BH are the same. Compared with the fiducial case, there are a few ($\sim 5$) BHs of above $v_{\rm cl}$ accreting at least one gas cell, for both the ``crucially bounded'' and ``unbounded'' group. We also note that above $\sqrt{2} v_{\rm cl}$, no BH accretes more than one gas cell, which is not runaway accretion.

From the cutoff of the CDF, we see that the fraction of BH seed with accretion events in the ``critically bounded'' and ``unbounded'' groups are lower, but still in the same order of magnitude. The maximum mass accretion for the three tests are also similar: although the cutoff of the orange line is higher, the general trend of the three lines at the high-mass end are very close if excluding the ``lucky'' BH.

In conclusion, we find that most of BH accretion events happens when $v_{\rm ini} \le v_{\rm cl}$. Our BH population in the main text overestimated these events, but still well within the same order of magnitude.

\begin{figure}
    \centering
    \includegraphics[width=\linewidth]{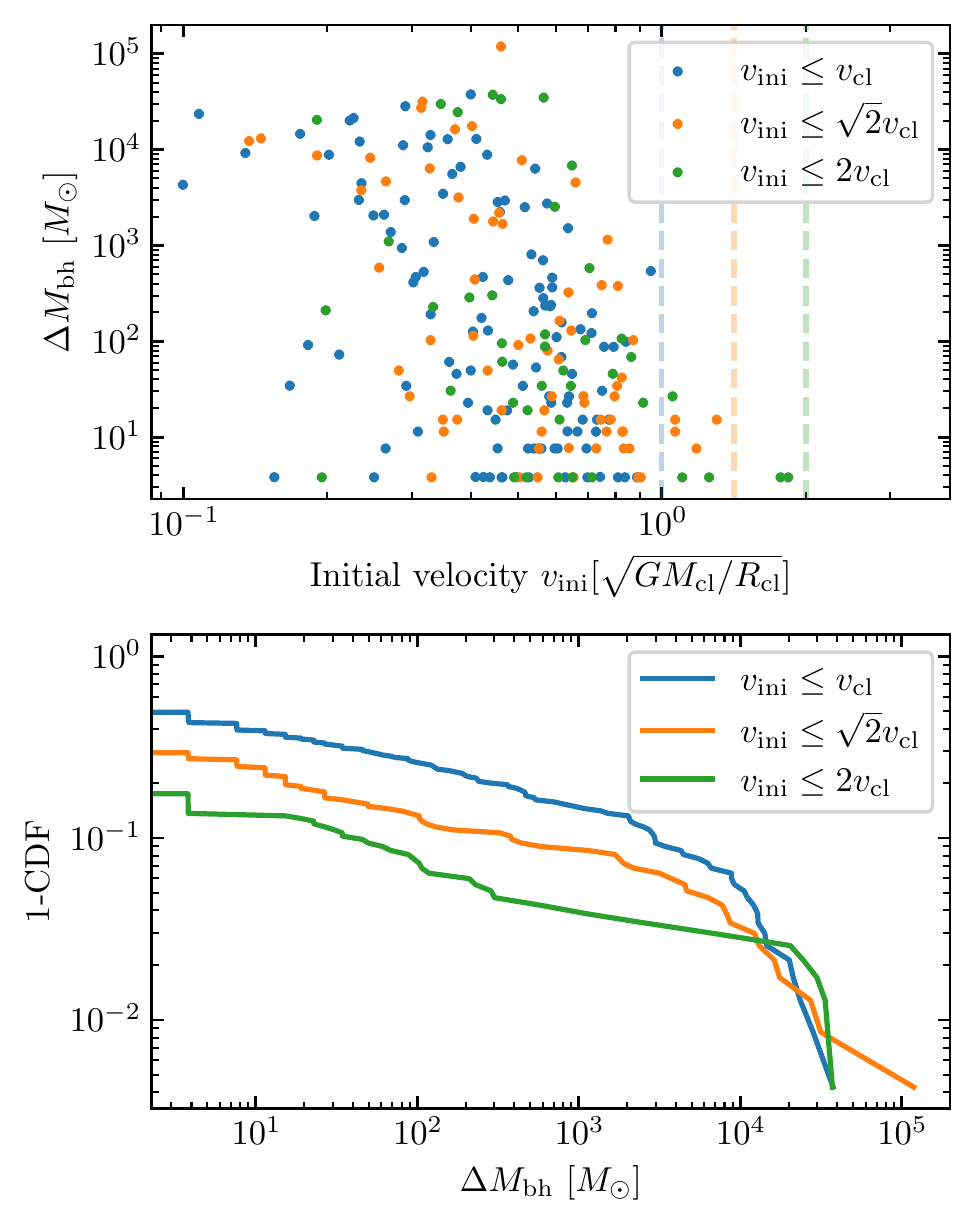}
    \vspace{-20pt}
    \caption{
    {A test of the dependence on the initial velocity distribution of BHs. In the fiducial case we confine $v_{\rm ini} \le v_{\rm cl}$, where $v_{\rm cl}=\sqrt{G M_{\rm cl}/R_{\rm cl}}$, and there are tests with different cutoffs ($\sqrt{2} v_{\rm cl}$ and $2 v_{\rm cl}$). } 
    }
    \label{fig:new_velocity}
\end{figure}

}

\end{appendix}

\end{document}